\documentclass[showpacs,preprintnumbers,amsmath,amssymb]{revtex4}
\usepackage{graphicx}
\usepackage{dcolumn}
\usepackage{amsmath}
\input{colordvi.tex}
 \def\apjl #1 #2 #3 {#1 Astrophys. J. {\bf #2} L#3}
 \def\mnras #1 #2 #3 {#1 Mon. Not. Roy. Astron. Soc. {\bf #2} #3}

\def\h#1{\hbox{${}^{#1}$H}}

\def\h502{\hbox{$ h^{2}_{50}$}}

\def\khat{\hat{k}}
\def\Khat{\hat{K}}

\def\la{\mathrel{\mathpalette\fun <}}
\def\ga{\mathrel{\mathpalette\fun >}}
\def\fun#1#2{\lower3.6pt\vbox{\baselineskip0pt\lineskip.9pt
  \ialign{$\mathsurround=0pt#1\hfil##\hfil$\crcr#2\crcr\sim\crcr}}}
\newcommand{\st}[2]{\stackrel{\mbox{\tiny (#1)}}{#2}\hspace{-0.1cm}}
\def\khat{\hat{k}}
\def\Khat{\hat{K}}

\begin{document}
%
\title{Magnetic Field Spectrum at Cosmological Recombination}
\author{%
Kiyotomo Ichiki$^{1}$ \footnote{E-mail address:
ichiki@resceu.s.u-tokyo.ac.jp},  
Keitaro Takahashi$^2$, 
Naoshi Sugiyama$^3$,
Hidekazu Hanayama$^4$,
Hiroshi Ohno$^5$
}
\affiliation{%
$^1$Research Center for the Early Universe, University of Tokyo, 7-3-1
Hongo, Bunkyo-ku, Tokyo 113-0033, Japan}
\affiliation{%
$^2$Department of Physics, Princeton University, Princeton, New Jersey
08544, USA}
\affiliation{%
$^3$Department of Physics and Astrophysics, Nagoya University, Nagoya
464-8602, Japan
}
\affiliation{%
$^4$National Astronomical Observatory of Japan, Tokyo 181-8588, Japan
}
\affiliation{%
$^5$Corporate Research and Development Center, Toshiba Corporation,
Kawasaki 212-8582, Japan
}
\date{\today}
\begin{abstract}
 A generation of magnetic fields from cosmological density
 perturbations is investigated.  In the primordial plasma before
 cosmological recombination, all of the materials except dark matter
 in the universe exist in the form of photons, electrons, and protons
 (and a small number of light elements). Due to the different
 scattering nature of photons off electrons and protons, electric
 currents and electric fields are inevitably induced, and thus
 magnetic fields are generated.  We present a detailed formalism of
 the generation of cosmological magnetic fields based on the framework
 of the well-established cosmological perturbation theory following
 our previous works.  We numerically obtain the power spectrum of
 magnetic fields for a wide range of scales, from $k\sim 10^{-5}$
 Mpc$^{-1}$ up to $k\sim 2\times 10^{4}$ Mpc$^{-1}$ and provide its
 analytic interpretation.
 Implications of these cosmologically generated magnetic fields are
 discussed.
\end{abstract}
\maketitle

\section{Introduction}

There are observational evidences which indicate that magnetic fields
exist not only galaxies, but also in even larger systems, such as
cluster of galaxies and extra-cluster spaces \cite{2002RvMP...74..775W}.
In addition to gravity, they would play an important role in the
formation processes of various objects and their dynamical evolution in
the universe. The effects of large scale magnetic fields on galaxy
formation and on formation of the first stars are extensively studied
\cite{2006astro.ph.11707G,2001A&A...378..777D,Dolag:2002bw,2006MNRAS.371..444S,2006ApJ...647L...1M}.

Yet, the origin of such large scale magnetic fields is still a mystery
\cite{2002RvMP...74..775W}.  It is now widely believed that the
magnetic fields at large scales are amplified from a tiny field and
maintained by the hydro-magnetic processes, i.e., the dynamo.
However, the dynamo needs a seed field to act on and does not explain
the origin of magnetic fields. As far as the magnetic fields in
galaxies are concerned, the seed fields as large as $10^{-20 \sim
-30}$ G are required in order to account for the observed fields
\cite{1999PhRvD..60b1301D} of order $1 \mu$ G at the present universe.
Discoveries of magnetic fields at damped Ly$\alpha$ systems
\cite{1992ApJ...388...17W} in the early universe and an intervening
galaxy at intermediate redshift \cite{1992ApJ...387..528K} may require
even larger seed fields.

The question arises here is, thus, what is the origin of such seed
fields?  Most of previous theories to explain the origin of seed
fields can be categorized into two, i.e., astrophysical and
cosmological mechanisms. Almost all astrophysical mechanisms of
generating seed magnetic fields are based on the Biermann battery
effect \cite{Biermann50}. This mechanism is operative when the
gradients in thermodynamic quantities, such as temperature and
density, are not parallel to that of pressure.  This mechanism has
been applied to various astrophysical systems, which include stars
\cite{1982PASP...94..627K}, supernova remnants
\cite{1998MNRAS.301..547M,Hanayama05}, protogalaxies
\cite{2000ApJ...540..755D}, large-scale structure formation
\cite{1997ApJ...480..481K}, and ionization front at cosmological
recombination \cite{2000ApJ...539..505G}.  These studies show that
magnetic fields with amplitude $10^{-16}$ $\sim$ $10^{-21}$ G could be
generated. Recently, it is proposed that the Weibel instability at the
structure formation shocks can generate rather strong magnetic fields
$10^{-7}$ G \cite{2005MNRAS.364..247F,Medvedev:2005ep}.  However, the
coherence-length of seed fields generated by astrophysical mechanisms,
especially the Weibel instability, tends to be too small to account
for galaxy-scale magnetic fields.

On the other hand, cosmological mechanisms based on inflation can
produce magnetic fields with a large coherence length since
accelerating expansion during inflation can stretche small-scale
fields to scales that can exceed the causal horizon. However, in the
simplest models with the usual electromagnetic field, which is
conformally coupled to gravity, the energy density in a fluctuation of
the field diminishes as $a^{-4}$ (where $a$ is cosmic scale factor) to
lead a negligible amplitude of magnetic fields at the end of
inflation. Therefore, to create enough amount of magnetic fields which
survive an expansion of the universe until today, some exotic
couplings between the electromagnetic field to other fields, such as
dilaton \cite{Ratra:1991bn,Bamba:2003av}, Higgs type scalar particles
\cite{2004PhRvD..70d3004P}, or gravity \cite{Turner:1987bw}, must be
introduced. Although some models of them are considered to be natural
extensions of standard particle model, the nature of generated
magnetic fields depends highly on the way of extension, whose validity
can never be tested experimentally by terrestrial particle
accelerators.  Moreover, it is recently argued that almost all models
that generate magnetic fields during inflation at the galactic scale
($\lambda \sim 0.1$ Mpc) are severely constrained as $B_\lambda \la
10^{-39}$ G \cite{Caprini:2001nb,2005PhRvD..72h8301C}, for a blue
primordial spectrum.  This is because anisotropic stress of magnetic
fields would produce a large amount of gravitational waves, which then
spoil the standard Big Bang nucleosynthesis by bringing an
overproduction of helium nuclei (and deuterium). 

In addition to the two categories described above, however, there is a
third category for the generation of large-scale seed fields:
cosmological fluctuations in the universe can create magnetic fields
prior to cosmological recombination. This category dates back to
Harrison (1970) \cite{1970MNRAS.147..279H} and has been rigorously
studied in recent years.  The attractive point is that mechanisms
based on cosmological perturbations are much less ambiguous than the
previous two categories because cosmological pertrubations have now
been well understood both theoretically and observationally through
cosmic microwave background (CMB) anisotropy and large scale
structure of the universe.  Thus it is possible to make a robust
quantitative evaluation of the generated magnetic fields.

Originally, Harrison found that the vorticity in a primordial plasma
can generate magnetic fields. This is because electrons and ions would
tend to spin at different rates as the universe expands due to the
radiation drag on electrons, arising rotation-type electric current
and thus inducing magnetic fields.  It was found that magnetic fields
of $10^{-21}$ G would result if the vorticity is equivalent to
galactic rotation by $z=10$. More recently, magnetic field generation
at recombination was investigated by evaluating the induced electric
field which arise in the matter flow dragged by a dipole photon field,
resulting in a seed field $10^{-26}$ G \cite{Hogan:2000gv}. Following
Harrison's idea, Matarrese reported that $10^{-29}$ G seed fields can
arise from the vorticity generated by second-order density
perturbations at recombination \cite{2005PhRvD..71d3502M}. In a
similar analysis, a larger value, $10^{-21}$ G, was obtained by
considering earlier periods when the energy density of photons was
larger and the photon mean free path was smaller
\cite{2004APh....21...59B}.  Other specific second order effects from 
the coupling between density and velocity fluctuations on the seed
field generation was evaluated in \cite{2005MNRAS.363..521G}.

In \cite{2005PhRvL..95l1301T}, we presented a formalism to calculate
the spectrum of seed fields based on the cosmological perturbation
theory.  There are three essential points in order to obtain the
correct spectrum.  First of all, we have to treat electron, proton and
photon fluids separately to evaluate the amount of electric current
and electric field. Secondly, we need a precise evaluation of
collision terms between three fluids, which generate the difference in
motion between them.  At this point, we found that not only the
velocity difference between electron and photon fluids, which is
conventionally studied, but also photon anisotropic stress contributes
to the difference in motion between electron and proton
fluids. Finally, the second-order perturbation theory is necessary
because vector-type perturbations, such as vorticity and magnetic
field, are known to be absent at the first order.  Based on this
formalism, in \cite{2006Sci...311..827I}, we calculated the spectrum
of generated magnetic fields at a range between $k \sim 10^{-5}$
Mpc$^{-1}$ and $k \sim 10$ Mpc$^{-1}$. Then we showed that the
magnetic field is contributed mainly from the velocity difference
between electrons and photons, which we called the "baryon-photon slip
term", on large scales ($k \lesssim 1$ Mpc$^{-1}$), while photon
anisotropic stress is dominant on small scalles ($k \gtrsim 1$
Mpc$^{-1}$).  As a whole, the spectrum is small-scale-dominant and the
amplitude is about $10^{-20}$ G at $k = 1$ Mpc$^{-1}$.

In this paper, we present details of our formalism and extend it in
several directions.  First, in section II, we describe the derivation
of equations of motion for electron and proton fluids, carefully
evaluating the collision term between electrons and photons, and then
obtain the evolution equation for magnetic fields. We calculate and
interpret the resulting spectrum numerically in section III. In
section IV, we derive the analytical expression of the spectrum, which
allows us to confirm the numerical results and extrapolate the
spectrum into much smaller scales than can be obtained numerically.
In section V, we discuss about the magnetic felicity, which in fact 
explicitly vanishes.  
Finally, in section VI, we give some discussions and conclusions.

\section{Basic Equations}

Here we derive basic equations for the generation of magnetic fields,
i.e.,  perturbation equations of photon, proton and
electron fluids.  While protons and electrons are conventionally treated
as a single fluid, however, it is necessary to deal with proton and
electron fluids separately in order to discuss the generation of
magnetic fields.  Let us begin with the Euler equations. Those are given
by
\begin{eqnarray}
&& m_{p} n u_{p}^{\mu} u_{pi;\mu} - e n u_{p}^{\mu} F_{i\mu}
   = C^{pe}_{~i} + C^{p \gamma}_{~i},
\label{eq:EOM_p2_0} \\
&& m_{e} n u_{e}^{\mu} u_{ei;\mu} + e n u_{e}^{\mu} F_{i\mu} =
C^{ep}_{~i} + C^{e \gamma}_{~i}~,
\label{eq:EOM_e2_0}
\end{eqnarray}
where $m_{p(e)}$ is the proton (electron) mass, $u_{p(e)}$ is the bulk
velocity of protons (electrons), $F_{\mu i}$ is the usual Maxwell
tensor.  The thermal pressure of proton and electron fluids are
neglected.  Here $\mu, \nu = 0,1,2,3$ and $i = 1,2,3$.  The r.h.s. of
Eq. (\ref{eq:EOM_p2_0}) and (\ref{eq:EOM_e2_0}) represent the
collision terms. The first terms in Eqs. (\ref{eq:EOM_p2_0}) and
(\ref{eq:EOM_e2_0}) are collision terms for the Coulomb scattering
between protons and electrons, which is given by
\begin{equation}
C^{pe}_{i} = - C^{ep}_i = -(u_{p i} - u_{e i}) e^2 n^2 \eta~,
\end{equation}
where 
\begin{equation}
\eta = \frac{\pi e^2 m_e^{1/2}}{(k_B  T_e)^{3/2}} \ln \Lambda
\sim 9.4 \times 10^{-16} {\rm sec} \left(\frac{1+z}{10^5}\right)^{-3/2}
     \left(\frac{\ln\Lambda}{10}\right)~,
\end{equation}
is the resistivity of the plasma and $\ln\Lambda$ is the Coulomb logarithm.
As is well known, this term acts as the diffusion term in the evolution equation
of magnetic field. The importance of the diffusion effect can be estimated
by the diffusion scale,
\begin{equation}
\lambda_{\rm diff}
\equiv \sqrt{\eta \tau}
\sim 100 \left(\frac{\tau}{H_{0}^{-1}}\right)^{1/2} {\rm AU},
\label{eq:diffusion}
\end{equation}
above which magnetic field cannot diffuse in the time-scale $\tau$.
Here $H_{0} = 70 {\rm km/s/Mpc}$ is the present Hubble parameter.
Thus, at cosmological scales considered in this paper, this term can be
safely neglected.

The other terms expressed by $C^{p(e) \gamma}_{~i}$ are the collision
terms for Compton scattering of protons (electrons) with photons.
Since photons scatter off electrons preferentially compared with protons
by a factor of $(m_e/m_p)^2$, we can safely drop the term $C^{p \gamma}_{~i}$
from the Euler equation of protons. This difference in collision terms
between protons and electrons ensures that small difference in velocity
between protons and electrons, that is, electric current, is indeed generated
once the Compton scattering becomes effective. 

\subsection{Compton Collision Term}
Let us now evaluate the Compton scattering term. In the limit of
completely elastic collisions between photons and electrons, this term
vanishes. Typically, in the regime of interest in this paper, very
little energy is transfered between electrons and photons in Compton
scatterings.  Therefore it is a good approximation to expand the
collision term systematically in powers of the energy transfer.

Let us demonstrate this specifically. We consider the collision process
\begin{equation}
\gamma(p_{i}) + e^{-}(q_{i}) \rightarrow \gamma(p'_{i}) + e^{-}(q'_{i}),
\end{equation}
where the quantities in the parentheses denote the particle momenta.
To calculate this process, we evaluate the collision term in the
Boltzmann equation of photons:
\begin{eqnarray}
C[f(\vec{p})]&=&\frac{a}{p}\int\frac{d^3q}{(2\pi)^3 2E_e(q)}
 \int\frac{d^3q^\prime}{(2\pi)^3 2E_e(q^\prime)}
 \int\frac{d^3p^\prime}{(2\pi)^3 2E(p^\prime)}(2\pi)^4|M|^2 \nonumber \\
 &&\times \delta^{(3)}[\vec{p}+\vec{q}-\vec{p^\prime}-\vec{q^\prime}]
  \delta[E(p)+E_e(q)-E(p^\prime)-E_e(q^\prime)] \nonumber \\
 &&\times [f_e(\vec{q^\prime})f(\vec{p^\prime})(1+f(\vec{p}))-f_e(\vec{q})f(\vec{p})(1+f(\vec{p^\prime}))]~,
\end{eqnarray}
where $f(\vec{p})$ and $f_e(\vec{q})$ are the distribution functions
of photons and electrons, $E_e(q) = \sqrt{q^2+m_e^2}$ is the energy of
an electron, and the delta functions enforce the energy and momentum
conservations.  We have dropped the Pauli blocking factor $(1-f_e)$.
The Pauli blocking factor can be always omitted safely in the epoch of
interest, because $f_e$ is very small after electron-positron
annihilations.  Note that the stimulating factor can be also
dropped because this does not contribute to the Euler equation.  

Integrating over $\vec{q^\prime}$, we obtain
\begin{eqnarray}
C[f(\vec{p})]&=&\frac{a}{p}\int\frac{d^3q}{(2\pi)^3 2E_e(q)}
 \int\frac{d^3p^\prime}{(2\pi)^3
 2E(p^\prime)}\frac{(2\pi)}{2E_e(|\vec{q}+\vec{p}-\vec{p^\prime}|)}|M|^2
 \nonumber \\
 &&\times \delta[E(p)+E_e(q)-E(p^\prime)-E_e(|\vec{q}+\vec{p}-\vec{p^\prime}|)] \nonumber \\
 &&\times [f_e(\vec{q}+\vec{p}-\vec{p^\prime})f(\vec{p^\prime})-f_e(\vec{q})f(\vec{p})]~.
\end{eqnarray}
In the regime of our interest, energy transfer through the Compton
scattering is small and can be ignored in the first order density
perturbations. As we already discussed earlier, however,  
it is essential to take the second order couplings in the
Compton scattering term into consideration for generation of magnetic 
fields. Therefore we expand the collision term up to the first order in 
powers of the  energy transfer$^\dag$, and
keep terms up to second order in density perturbations. 
\footnotetext[2]{However, we shall keep up to the second order terms
for the purpose of reference.}

The expansion parameter is the energy transfer,
\begin{equation}
E_e(q)-E_e(q^\prime)=\frac{q^2}{2m_e}
 -\frac{(\vec{q}+\vec{p}-\vec{p^\prime})^2}{2m_e}
 \approx\frac{(\vec{p^\prime}-\vec{p})\cdot \vec{q}}{m_e}-\frac{(\vec{p}-\vec{p^\prime})^2}{2m_e}~,
\label{eq:E_difference}
\end{equation}
over the temperature of the universe.  Employing $p \sim T$, we can
estimate the order of this expansion parameter as
${\cal O}(\frac{pq}{m_e T}) \sim {\cal O}(\frac{q}{m_e})$, which is small
when electrons are non-relativistic.  Note that, in the cosmological
Thomson regime, electrons in the thermal bath of photons are
non-relativistic, $p \sim \frac{q^2}{2m_e}$, and the energy of
photons is much smaller than the rest mass of a electron, $p \ll m_e$.
Thus, it also holds that $q \sim \sqrt{2 m_e p} \gg p$, and the
second term in Eq.(\ref{eq:E_difference}) is usually smaller than the
first one.

Now let us divide the collision integral into four parts, i.e., the
denominators of the Lorentz volume, the scattering amplitude, the
delta function and the distribution functions, and expand them due to
the expansion parameter defined above.  First of all, the denominator
in the Lorentz invariant volume can be expanded to
\begin{eqnarray}
\frac{1}{E_e(q)E_e(|\vec{q}+\vec{p}-\vec{p^\prime}|)}
&=&\left(m_e+\frac{1}{2m_e}|\vec{q}+\vec{p}-\vec{p^\prime}|^2\right)^{-1}
 \left(m_e+\frac{1}{2m_e}|\vec{q}|^2\right)^{-1} \nonumber \\
&\approx&\frac{1}{m_e^2}\left(1-{\cal E}_{(\frac{q}{m_e})^2}
-{\cal E}_{(\frac{pq}{m_e^2})}-{\cal E}_{(\frac{p}{m_e})^2}\right) ,
\label{eq:11}
\end{eqnarray}
where
\begin{equation}
{\cal E}_{(\frac{q}{m_e})^2} = \frac{q^2}{m_e^2},~~~
{\cal E}_{(\frac{qp}{m_e^2})}
= \frac{(\vec{p}-\vec{p^\prime}) \cdot \vec{q}}{m_e^2},~~~
{\cal E}_{(\frac{p}{m_e})^2}
= \frac{(\vec{p}-\vec{p^\prime})^2}{2 m_e^2} .
\end{equation}

Secondly, we consider the matrix element.
The matrix element for Compton scattering in the rest frame of the
electron is given by,
\begin{eqnarray}
|M|^2 &=& 2(4\pi)^2 \alpha^2
 \left[\frac{\tilde{p^\prime}}{\tilde{p}}
  +\frac{\tilde{p}}{\tilde{p^\prime}}-\sin^2\tilde{\beta}\right]~,
 \nonumber \\
 \cos\tilde{\beta} &=& \tilde{\hat{p}}\cdot\tilde{\hat{p}}^\prime~,
\end{eqnarray}
where $\tilde{p}$ and $\tilde{p^\prime}$ are the energies of incident
and scattered photons, $\tilde{\hat{p}}$ and $\tilde{\hat{p}}^\prime$
are the unit vectors of $\tilde{\vec{p}}$ and
$\tilde{\vec{p}}^\prime$, respectively, denoting the directions of the
photons in this frame. The Lorentz transformation with electron's
velocity ($q/m_e$) gives the following relations,
\begin{eqnarray}
\frac{p}{\tilde{p}}&=&\frac{\sqrt{1-(q/m_e)^2}}{1-\vec{p}
 \cdot{\vec{q}} / pm_e} ,\\
 p_\mu p^\mu &=& \tilde{p_\mu}\tilde{p^\mu} .
\end{eqnarray}
Using these relations, we evaluate the matrix element in the CMB frame
as \cite{1995PhDT..........H}
\begin{eqnarray}
|M|^2&=&2(4\pi)^2\alpha^2 \left(
			  {\cal M}_0 
			 +{\cal M}_{\frac{q}{m_e}}
			 +{\cal M}_{(\frac{q}{m_e})^2}
			 +{\cal M}_{(\frac{qp}{m_e^2})}
			 +{\cal M}_{(\frac{p}{m_e})^2}
\right)~, \label{eq:15}\\
{\cal M}_0 &=& 1+\cos^2\beta~, \\
{\cal M}_{\frac{q}{m_e}} &=&
 -2\cos\beta(1-\cos\beta)\left[\frac{\vec{q}}{m_e}\cdot\left(\hat{p}+\hat{p^\prime}\right)\right]~, \\
{\cal M}_{(\frac{q}{m_e})^2}&=&
 2\cos\beta(1-\cos\beta)\frac{q^2}{m_e^2}~, \\
{\cal M}_{\frac{qp}{m_e^2}}&=&
 (1-\cos\beta)(1-3\cos\beta)\left[\frac{\vec{q}}{m_e}\cdot(\hat{p}+\hat{p^\prime})\right]^2+2\cos\beta(1-\cos\beta)\frac{(\vec{q}\cdot\hat{p})(\vec{q}\cdot\hat{p^\prime})}{m_e^2}~,\\
{\cal M}_{(\frac{p}{m_e})^2}&=&(1-\cos\beta)^2\frac{p^2}{m_e^2}~.
\end{eqnarray}

Thirdly, we expand the delta function to  
\begin{eqnarray}
&& \delta[p-p^\prime+E_e(q)-E_e(q^\prime)] \nonumber \\
&& \approx \delta(p-p^\prime)
   - \frac{\partial \delta (p-p^\prime+E_e(q)-E_e(q^\prime))}
          {\partial E_e(q^\prime)}|_{E_e(q)=E_e(q^\prime)}
     \left(E_e(q)-E_e(q^\prime)\right)
   \nonumber \\
&& ~~~ + \frac{1}{2} \frac{\partial^2 \delta (p-p^\prime + E_e(q)-E_e(q^\prime))}
                          {\partial p^2}|_{E_e(q)=E_e(q^\prime)}
         \left(E_e(q)-E_e(q^\prime)\right)^2 \nonumber \\
&& = \delta(p-p^\prime)
     + \frac{\partial \delta(p-p^\prime)}{\partial p^\prime}
       \left[\frac{(\vec{p}-\vec{p^\prime}) \cdot \vec{q}}{m_e} + \frac{(p-p^\prime)^2}{2m_e}\right]
     + \frac{1}{2} \left[\frac{(\vec{p}-\vec{p^\prime}) \cdot \vec{q}}{m_e}\right]^2
       \frac{\partial^2 \delta (p-p^\prime)}{\partial {p^\prime}^2} \nonumber \\
&& \equiv \delta(p-p^\prime)
          + \frac{\partial \delta(p-p^\prime)}{\partial p^\prime}
            \left[{\cal D}_{(\frac{q}{m})} + {\cal D}_{(\frac{p}{m})}\right]
          + \frac{1}{2} {\cal D}^2_{(\frac{q}{m})}
            \frac{\partial^2 \delta (p-p^\prime)}{\partial {p^\prime}^2} .
\end{eqnarray}
where
\begin{equation}
{\cal D}_{(\frac{q}{m})} \equiv \frac{(\vec{p}-\vec{p^\prime}) \cdot \vec{q}}{m_e}, ~~~
{\cal D}_{(\frac{p}{m})} \equiv \frac{(p-p^\prime)^2}{2m_e}.
\end{equation}

Finally, the distribution of the electron can be expanded to
\begin{equation}
f_e(\vec{q}+\vec{p}-\vec{p^\prime})\approx f_e(\vec{q})+\frac{\partial
 f_e}{\partial \vec{q}}\cdot
 (\vec{p}-\vec{p^\prime})+(p^i-{p^\prime}^i)\frac{\partial^2
 f_e}{\partial q^i \partial q^j}(p^j-{p^\prime}^j) .
\label{eq:dist_of_e}
\end{equation}
We assume that the electrons are kept in thermal equilibrium and in the
Boltzmann distribution:
\begin{equation}
f_e(\vec{q})=n_e \left(\frac{2\pi}{m_eT_e}\right)^{3/2}\exp
\left[-\frac{(\vec{q}-m_e\vec{v_e})^2}{2m_e T_e}\right]~,
\end{equation}
where $v_e$ is the bulk velocity of electrons. The derivatives of
the distribution function with respect to the momentum are given as
\begin{eqnarray}
\frac{\partial f_e}{\partial \vec{q}}&=&
 -f_e(\vec{q})\frac{\vec{q}-m_e\vec{v_e}}{m_e T_e}~,\\
\frac{\partial^2 f_e}{\partial q_i \partial q_j}&=&-\frac{\partial
 f_e(\vec{q})}{\partial q_j}\frac{q^i-m_e v^i}{m_e
 T_e}-f_e(\vec{q})\frac{\delta_{ij}}{m_e T_e}~.
\end{eqnarray}
By substituting above equations, Eq.(\ref{eq:dist_of_e}) is written as
\begin{eqnarray}
f_e(\vec{q}+\vec{p}-\vec{p^\prime})&\approx&
 f_e(\vec{q})-f_e(\vec{q})\frac{\vec{q}-m_e \vec{v_e}}{m_e
 T_e}\cdot(\vec{p}-\vec{p^\prime})+\frac{1}{2}f_e(\vec{q})\left[\left[\frac{(\vec{p}-\vec{p^\prime})(\vec{q}-m_e\vec{v_e})}{m_e
								 T_e}\right]^2-\frac{|\vec{p}-\vec{p^\prime}|^2}{m_e T_e}\right]\nonumber \\
&\equiv&f_e(\vec{q})\left[1-{\cal F}_{(\frac{q}{m})}+\frac{1}{2}{\cal
		     F}_{(\frac{q}{m})}^2-{\cal F}_{(\frac{p}{m})}\right]~.
\end{eqnarray}
Therefore, we have
\begin{equation}
f_e(\vec{q}+\vec{p}-\vec{p^\prime})f(\vec{p^\prime})-f_e(\vec{q})f(\vec{p})=
f_e(\vec{q})\left[f(\vec{p^\prime})-f(\vec{p})\right]
-f(\vec{p^\prime})f_e(\vec{q}){\cal F}_{(\frac{q}{m})}
-f(\vec{p^\prime})f_e(\vec{q})
\left[
{\cal F}_{(\frac{T}{m})}-\frac{1}{2}{\cal F}^2_{(\frac{q}{m})}
\right]~.
\end{equation}

Fortunately, it has been known that the leading term (zeroth
order term), obtained by multiplying together the first term in the
delta function and the zeroth order distribution functions, is
zero. It means that we only have to keep up to the first order terms
when we expand the matrix element and the energies, 
in order to keep the collision term up to the second order
\cite{1995ApJ...439..503D}. 

Therefore we have,
\begin{equation}
\frac{|M|^2}{E_e(q)E_e(q^\prime)}\approx6\pi\sigma_T({\cal M}_0+{\cal
M}_{\frac{q}{m_e}})
\end{equation}
Combining altogether, we obtain the collision term expanded with respect
to the energy transfer as (note that this expansion is
not with respect to the density perturbations)
\begin{equation}
 C[f]=\frac{a}{p}\left(\frac{\pi}{4}\right)\int \frac{d^3
  p^\prime}{(2\pi)^3 p^\prime}\frac{d^3 q}{(2\pi)^3}
\left[{\mbox{(0th order term)}}+{\mbox{(1st order terms)}}+{\mbox{(2nd order terms)}}\right]+\cdot\cdot\cdot
\end{equation}
where
\begin{eqnarray}
\mbox{0th order term:}&& \nonumber\\
&& 6\pi\sigma_T{\cal M}_0 \delta
 (p-p^\prime)f_e(\vec{q})\left[f(\vec{p^\prime})-f(\vec{p}))\right] ,\\
\mbox{1st order terms:}&& \nonumber\\
&& 6\pi\sigma_T{\cal M}_0 f_e(\vec{q})\left[
-\delta(p-p^\prime)f(\vec{p^\prime}){\cal F}_{\frac{q}{m}}
+\frac{\partial \delta (p-p^\prime)}{\partial p^\prime}
{\cal D}_{\frac{q}{m}}
\left[f(\vec{p^\prime})-f(\vec{p})\right]
\right]\nonumber \\
&&+6\pi\sigma_T{\cal M}_{\frac{q}{m}}f_e(\vec{q})\delta
 (p-p^\prime)\left[f(\vec{p^\prime})-f(\vec{p})\right] ,\\
\mbox{2nd order terms:}&& \nonumber\\
 && 6\pi\sigma_T {\cal M}_0 f_e(\vec{q})\Biggl[
-\delta (p-p^\prime)f(\vec{p^\prime})\left[
{\cal F}_{\frac{p}{m}}-\frac{1}{2}{\cal F}^2_{\frac{q}{m}}
\right]
+\frac{1}{2}\frac{\partial^2 \delta (p-p^\prime)}{\partial {p^\prime}^2}
{\cal D}^2_{\frac{q}{m}}\left[
f(\vec{p^\prime})-f(\vec{p})
\right]\nonumber \\
&&+\frac{\partial \delta (p-p^\prime)}{\partial p^\prime}
{\cal D}_{\frac{p}{m}}
\left[f(\vec{p^\prime})-f(\vec{p})\right]-\frac{\partial \delta (p-p^\prime)}{\partial p^\prime}\left(
{\cal D}_{\frac{q}{m}}+{\cal D}_{\frac{p}{m}}
\right)f(\vec{p^\prime}){\cal F}_{\frac{q}{m}}
\Biggr] \nonumber \\
&&+ 6\pi\sigma_T{\cal M}_{\frac{q}{m}}f_e(\vec{q})\left[
-\delta (p-p^\prime)f(\vec{p^\prime}){\cal F}_{\frac{q}{m}}
+\frac{\partial \delta (p-p^\prime)}{\partial p^\prime}\left(
{\cal D}_{\frac{q}{m}}+{\cal D}_{\frac{p}{m}}\left[
f(\vec{p^\prime}-f(\vec{p}))
\right]
\right)
\right] .
\end{eqnarray}
From now on, we omit the second order terms.  These terms are 
not only much smaller than the first order terms but also 
may not contribute to the Euler equation at all (see
\cite{2006astro.ph..4416B}). 
Evaluating the first moment of the
above collision term, we obtain the Compton scattering term in the
Euler equation (\ref{eq:EOM_e2}) as
\begin{eqnarray}
C^{e \gamma}_{i}&=&-\int \frac{d^3 p}{(2\pi)^3}p_i C[f] \nonumber \\
 &=&-\frac{4\sigma_T \rho_\gamma a n_e}{3}\left[(v_{e i} - v_{\gamma
 i})+\frac{1}{4}v_{ej} \Pi^{\phantom{\gamma}j}_{\gamma i} \right] .
\label{eq:coll}
\end{eqnarray}
Here moments of the distribution functions are given by
\begin{eqnarray}
&& \int \frac{d^{3}p}{(2\pi)^{3}} p f_{\gamma}(\vec{p}) = \rho_{\gamma}, \\
&& \int \frac{d^{3}p}{(2\pi)^{3}} p_{i} f_{\gamma}(\vec{p})
   = \frac{4}{3} \rho_{\gamma} v_{\gamma i}, \\
&& \int \frac{d^{3}q}{(2\pi)^{3}} f_{e}(\vec{q})
   = n_e, \\
&& \int \frac{d^{3}q}{(2\pi)^{3}} q_{i} f_{e}(\vec{q})
   = \rho_{e} v_{e i}, \\
&& \int \frac{d^{3}p}{(2\pi)^{3}} p^{-1} p_{i} p_{j} f_{\gamma}(\vec{p})
   = \frac{1}{3} \rho_{\gamma} \Pi_{\gamma ij} + \frac{1}{3} \rho_{\gamma} \delta_{ij},
\end{eqnarray}
where $\rho_\gamma$ and $\rho_e(=m_e n_e)$ are energy densities of
photons and electrons, $v^i_\gamma$ and $v^i_e$ are their bulk three velocities
defined by $v^i \equiv u^i / u^0$, and $\Pi^{ij}_\gamma$ is anisotropic stress
of photons. It should be noted that the collision term (\ref{eq:coll}) was obtained
nonperturbatively with respect to density perturbations
\cite{2005PhRvL..95l1301T}.

\subsection{Evolution Equations of Magnetic Fields}
Now we obtain the Euler equations for protons and electrons as
\begin{eqnarray}
&& m_{p} n u_{p}^{\mu} u_{p i;\mu} - e n u_{p}^{\mu} F_{i\mu}
   = 0,
\label{eq:EOM_p2} \\
&& m_{e} n u_{e}^{\mu} u_{e i;\mu} + e n u_{e}^{\mu} F_{i\mu}
   = - \frac{4 \sigma_{T} \rho_{\gamma} a n}{3}
       \left[ ( v_{e i} - v_{\gamma i}) + \frac{1}{4} v_{ej} \Pi_{\gamma
	i}^{\phantom{\gamma}j} \right],
\label{eq:EOM_e2}
\end{eqnarray}
where $m_{p}$ is the proton mass. Here we ignore the pressure of
proton and electron fluids. Also the Coulomb collision term is
neglected as explained below Eq. (\ref{eq:diffusion}). Note that the
collision term was not evaluated in a manifestly covariant way. Here
the left hand side in Eqs. (\ref{eq:EOM_p2}) and (\ref{eq:EOM_e2})
should be evaluated in conformal coordinate system. We also assumed
the local charge neutrality: $n = n_{e} \sim n_{p}$.  In the case
without electromagnetic fields ($F_{i\mu}=0$), the sum of the
equations (\ref{eq:EOM_p2}) and (\ref{eq:EOM_e2}) gives the Euler
equation for the baryons in the standard perturbation theory.  On the
other hand, subtracting Eq. (\ref{eq:EOM_p2}) multiplied by $m_{e}$
from Eq. (\ref{eq:EOM_e2}) multiplied by $m_{p}$, we obtain
\begin{eqnarray}
&& - \frac{m_{p}m_{e}}{e}
   \left[ n u^{\mu} \left( \frac{j_{i}}{n} \right)_{;\mu}
          + j^{\mu}
            \left( \frac{m_{p} - m_{e}}{m_{p} + m_{e}} \frac{j_{i}}{en} - u_{i} \right)_{;\mu}
   \right]
 + e n ( m_{p} + m_{e} ) u^{\mu} F_{i\mu}
   - ( m_{p} - m_{e} ) j^{\mu} F_{i\mu} \nonumber \\
&& = - \frac{4 m_{p}  \rho_{\gamma} a n \sigma_{T}}{3}
       \left[ ( v_{e i} - v_{\gamma i} ) + \frac{1}{4} v_{ej}
	\Pi_{\gamma i}^{\phantom{\gamma}j} \right],
\label{eq:Ohm1}
\end{eqnarray}
where $u^{\mu}$ and $j^{\mu}$ are the center-of-mass 4-velocity of the proton and
electron fluids and the net electric current, respectively, defined as
\begin{eqnarray}
&& u^{\mu} \equiv \frac{m_{p}u_{p}^{\mu} + m_{e}u_{e}^{\mu}}{m_{p} + m_{e}}, \\
&& j^{\mu} \equiv e n (u_{p}^{\mu} - u_{e}^{\mu}).
\end{eqnarray}
Employing the Maxwell equations $F^{\mu\nu}_{;\nu} = j^{\mu}$, we see that
the quantities in the square bracket in the l.h.s. of Eq. (\ref{eq:Ohm1}) is
suppressed at the recombination epoch, compared to the second term,
by a factor \cite{1994MNRAS.271L..15S}
\begin{equation}
\frac{c^{2}}{L^{2} \omega_{p}^{2}}
\sim 3 \times 10^{-40}
     \left( \frac{10^{3} {\rm cm}^{-3}}{n} \right)
     \left( \frac{1 {\rm Mpc}}{L} \right)^{2},
\end{equation}
where $c$ is the speed of light, $L$ is a characteristic length of the system
and $\omega_{p} = \sqrt{4 \pi n e^{2}/m_{e}}$ is the plasma frequency.

The third term in the l.h.s. of Eq. (\ref{eq:Ohm1}), 
i.e., $(m_p-m_e)j^\mu F_{i\mu}$, is the Hall term
which can also be neglected because the Coulomb coupling between protons
and electrons is so tight that $|u^{i}| \gg |u_{p}^{i} -
u_{e}^{i}|$. Then we obtain a generalized Ohm's law: 
\begin{equation}
u^{\mu} F_{i\mu}
= - \frac{4 \sigma_{T} \rho_{\gamma} a}{3 e}
      \left[ ( v_{e i} - v_{\gamma i} ) + \frac{1}{4} v_{ej} \Pi_{\gamma
       i}^{\phantom{\gamma}j} \right]
\equiv C_{i}.
\end{equation}

Now we derive the evolution equation for the magnetic field, which 
can be obtained from the Bianchi identities $F_{[\mu\nu,\lambda]} =
0$, as
\begin{eqnarray}
0
&=& \frac{3}{2} \epsilon^{ijk} u^{\mu} F_{[jk,\mu]} \nonumber \\
&=& u^{\mu} {\cal B}^{i}_{~,\mu}
    - \epsilon^{ijk} \left( C_{j,k} + \frac{u^{0}_{~,j}}{u^{0}} C_{k} \right)
    - ( u^{i}_{~,j} {\cal B}^{j} - u^{j}_{~,j} {\cal B}^{i} )
    + \frac{u^{0}_{~,j}}{u^{0}} ( {\cal B}^{j} u^{i} - {\cal B}^{i} u^{j} ),
\label{eq:Bianchi}
\end{eqnarray}
where $\epsilon^{ijk}$ is the Levi-Civit\`{a} tensor and ${\cal B}^i
\equiv (a^2 B^{i}) = \epsilon^{ijk} F_{jk}/2$ is the magnetic field in
the comoving frame \cite{1998PhRvD..57.3264J}. We will now expand the
photon energy density, fluid velocities and photon anisotropic stress
with respect to the density perturbation as
\begin{eqnarray}
&& \rho_{\gamma}(t,x_i) = \st{0}{\rho}_{\gamma}(t) + \st{1}{\rho}_{\gamma}(t,x_i) + \cdots, ~~~
   u^{0}(t,x_i) = a(t)^{-1} + \st{1}{u}^{0}(t,x_i) + \cdots, \nonumber \\
&& u^{i}(t,x_i) = \st{1}{u}^{i}(t,x_i) + \st{2}{u}^{i}(t,x_i) + \cdots, ~~~
   v_{i}(t,x_i) = \st{1}{v}_{i}(t,x_i) + \st{2}{v}_{i}(t,x_i) + \cdots \nonumber \\
&& \Pi_{\gamma}^{ij}(t,x_i) = \st{1}{\Pi}_{\gamma}^{ij}(t,x_i) + \cdots,
\end{eqnarray}
where the superscripts $(0), (1)$, and $(2)$ denote the order of expansion and $t$
is the cosmic time. Remembering that ${\cal B}^{i}$ is a second-order quantity, we see that
all terms involving ${\cal B}^{i}$ in Eq. (\ref{eq:Bianchi}), other than the first term,
can be neglected. Thus we obtain
\begin{eqnarray}
\frac{d{\cal B}^{i}}{dt}
&\sim& \epsilon^{ijk} \left( C_{j,k} + \frac{u^{0}_{~,j}}{u^{0}} C_{k} \right) \nonumber \\
&=& \frac{4 \sigma_{T} \st{0}{\rho}_{\gamma} a}{3 e} \epsilon^{ijk}
    \Biggl[ \left( \st{1}{\delta_{\gamma,j}} - 2 \st{1}{\Phi}_{,j} \right)
            \left( \st{1}{v}_{ek} - \st{1}{v}_{\gamma k} \right)
	    - \frac{1}{4} \left( \st{1}{v}_{el}
			   \st{1}{\Pi}{}^{l}_{\gamma j} \right)_{,k} 
            - 2 \st{1}{h}^{l}_{~j,k}
              \left( \st{1}{v}_{el} - \st{1}{v}_{\gamma l} \right)
            - \left( \st{2}{v}_{ej,k} - \st{2}{v}_{\gamma j,k} \right)
    \Biggr],
\label{eq:B_dot}
\end{eqnarray}
where $\st{1}{\Phi}$ and $\st{1}{h}^{i}_{~j}$ are first-order curvature
and tensor 
perturbations, respectively, in Poisson gauge, and we used the density
contrast of photons, 
$\st{1}{\delta}_{\gamma,k} \equiv \st{1}{\rho}_{\gamma,k} /
\st{0}{\rho}_{\gamma}$. 
Further, we employed the fact that there is no vorticity in the linear order:
$\epsilon^{ijk} \st{1}{v}_{j,k} = 0$.  It should be noted that the velocity
of electron fluid can be approximated to the center-of-mass velocity at this order,
$\st{1}{v}_{e}^{i} \sim \st{1}{v}^{i}$.  
The physical meaning of this equation is that 
electrons gain (or lose) their momentum through scatterings due
to the relative velocity to photons (baryon-photon slip), and the
anisotropic pressure from photons. The momentum transfer from the
photons ensures the velocity difference between electrons and protons,
and thus eventually generates magnetic fields. 
We found that the contribution from the curvature perturbation is
always much smaller than that from  the density contrast of photons in
the first term in Eq.(\ref{eq:B_dot}) (which will be clearly seen in
Figs.\ref{fig:4} and \ref{fig:perturbation_spectra}). Therefore we
shall omit the curvature perturbation 
hereafter when considering the evolution of magnetic fields.

The first term in Eq. (\ref{eq:B_dot}) is exactly the same
discussed in \cite{2005MNRAS.363..521G}. They have estimated
contributions from these terms by considering typical values at
recombination. Here we solve the equation numerically and obtain 
a robust prediction of the amplitude of magnetic fields.

Eq.(\ref{eq:B_dot}) shows that the magnetic field cannot be generated
in the first (linear) order.  The r.h.s. of Eq.(\ref{eq:B_dot})
contains two kinds of source terms, i.e., intrinsic second order
quantities and products of first order quantities.  Since the first
order quantities can be exactly evaluated within the frame work of the
standard cosmological linear perturbation theory, we hereafter
concentrate on the products of first order quantities.  The evaluation
of the intrinsic second order term will be left for the future work.

\section{Cosmological Magnetic Fields}

\subsection{Spectrum of Cosmologically Generated Magnetic Fields}

In this section we derive the spectrum of magnetic fields generated by
cosmological perturbations. The cosmological perturbations can be
decomposed into scalar, vector, and tensor modes. These modes
physically correspond to perturbations of density, vorticity and
gravitational waves, respectively.  These cosmological perturbations
are most likely to be generated during the inflation epoch. Most of
the simple single field inflation models predict generation of
adiabatic scalar and tensor type perturbations but vector type
perturbations.  Even if the vector type perturbations are generated,
they are known to be damped away in the expanding universe.
Observations of CMB temperature fluctuations and large scale structure
of the universe strongly suggest the adiabatic scalar perturbations
while we only have upper limits on tensor
perturbations~\cite{2005PhRvD..71j3515S}.  Therefore at most tensor
perturbations are only sub-dominant components.  Hence we 
consider only scalar type perturbations for linear order quantities
throughout this paper.


The linear-order scalar perturbations can be written as 
\begin{eqnarray}
 \delta_{\gamma,k}&=&\frac{1}{\sqrt{(2\pi)^3}}\int(ik_k)\delta_\gamma(\vec{k},t)e^{i\vec{k}\cdot\vec{x}}d^3\vec{k}, \\
 v_{el}&=&\frac{1}{\sqrt{(2\pi)^3}}\int(-i\khat_l)v_e(\vec{k},t)
  e^{i\vec{k}\cdot\vec{x}}d^3 \vec{k}, \\
 v_{el,k}&=&\frac{1}{\sqrt{(2\pi)^3}}\int \khat_l k_k v_e(\vec{k},t)
  e^{i\vec{k}\cdot\vec{x}}d^3 \vec{k}, \\
  \Pi^l_j
  &=&\frac{1}{\sqrt{(2\pi)^3}}\int(-\khat^l\khat_j+\frac{1}{3}\delta^l_j)\Pi(\vec{k},t)
  e^{i\vec{k}\cdot\vec{x}}d^3 \vec{k}, \\
 \Pi^l_{j,k}
  &=&\frac{1}{\sqrt{(2\pi)^3}}\int(ik_k)(-\khat^l\khat_j+\frac{1}{3}\delta^l_j)\Pi(\vec{k},t)
  e^{i\vec{k}\cdot\vec{x}}d^3 \vec{k}.
\end{eqnarray}
Here $\khat^i \equiv k^i/k$. 
By using these expressions, we can write
the evolution equation of magnetic fields in Fourier space as,
\begin{equation}
(a^2 {B^i}(\vec{K},t))^\cdot=\frac{4\sigma_T\rho_\gamma}{3e}\epsilon^{ijk}
 \frac{1}{\sqrt{(2\pi)^3}}\int d^3\vec{k'}
\left[
 \khat '_j K_k \delta_\gamma(\vec{K}-\vec{k'},t)\delta v(\vec{k'},t)
-\frac{1}{4}\alpha_{jk}(\vec{K},\vec{k'})
v_e(\vec{K}-\vec{k'},t)\Pi(\vec{k'},t)
\right]
\label{eq:a2Bdot}
\end{equation}
where we have defined,
\begin{eqnarray}
\delta v &=& v_e - v_\gamma, \\ 
\alpha_{jk}(\vec{K},\vec{k^\prime}) &=&
\frac{K}{|\vec{K}-\vec{k^\prime}|}
\left(K(\Khat \cdot \khat^\prime) - \frac{4}{3}k^\prime\right)
\Khat_k \khat_j^\prime.
\end{eqnarray}
To obtain the spectrum, one needs to evaluate
$\left<B^i(\vec{K})B^\ast_i(\vec{K^\prime})\right> =
S(K)\delta(\vec{K}-\vec{K^\prime})$. That is
\begin{eqnarray}
a^4 B^i(\vec{K}) B^*_i(\vec{K'})
&=& \left(\frac{4\sigma_T}{3e}\right)^2
  \epsilon^{ijk}\epsilon^{lm}_i\displaystyle\frac{1}{(2\pi)^3}
  \int d^3\vec{k'} d^3\vec{k} \int_{0}^{t_0} dt' \int_{0}^{t_0} dt'' \nonumber \\
&\times& \biggl[\khat '_j K_k
  \delta_\gamma(\vec{K}-\vec{k'},t')\delta v(\vec{k'},t')a(t')\rho_\gamma(t')
  \khat_l K'_m \delta^*_\gamma(\vec{K'}-\vec{k},t'')\delta v^*(\vec{k},t'')
  a(t'')\rho_\gamma(t'') \nonumber \\
 &&-\frac{1}{4}\khat '_j K_k \delta_\gamma(\vec{K}-\vec{k'},t')
  \delta v(\vec{k'},t')a(t')\rho_\gamma(t')\alpha_{lm}(\vec{K'},\vec{k})
  v^*_e(\vec{K'}-\vec{k},t'')\Pi^*(\vec{k},t'')a(t'')\rho_\gamma(t'')
  \nonumber \\
&&-\frac{1}{4}\alpha_{lm}(\vec{K},\vec{k'})
v_e(\vec{K}-\vec{k'},t')\Pi(\vec{k'},t') a(t')\rho_\gamma(t')
\khat_l K'_m \delta^*_\gamma(\vec{K'}-\vec{k},t'')
  \delta v^*(\vec{k},t'')a(t'')\rho_\gamma(t'') \nonumber \\
&&+\frac{1}{16}\alpha_{jk}(\vec{K},\vec{k'})v_e(\vec{K}-\vec{k'},t')
 \Pi(\vec{k'},t')a(t')\rho_\gamma(t')\alpha_{lm}(\vec{K'},\vec{k})
 v_e^*(\vec{K'}-\vec{k},t'')\Pi^*(\vec{k},t'')a(t'')\rho_\gamma(t'')\biggl]~.
 \end{eqnarray}
The final task we should do is to take an ensemble average of this
expression. Assuming that perturbations are random Gaussian variables
and using the Wick's theorem, we can expand the ensemble average of the
products of four Gaussian variables as, for example, 
\begin{eqnarray}
 \left<\delta_\gamma(\vec{K}-\vec{k'},t')\delta v(\vec{k'},t')
 \delta_\gamma^*(\vec{K'}-\vec{k},t'')\delta v^*(\vec{k},t'')\right>
 &=& \left< \delta_\gamma(\vec{K}-\vec{k'},t')\delta v(\vec{k'},t') \right>
 \left< \delta_\gamma^*(\vec{K'}-\vec{k},t'')\delta
 v^*(\vec{k},t'')\right>\nonumber \\
&+&
 \left<\delta_\gamma(\vec{K}-\vec{k'},t')\delta_\gamma^*(\vec{K'}-\vec{k},t'')
 \right> \left<\delta v(\vec{k'},t')\delta v^*(\vec{k},t'') \right>
 \nonumber \\
&+&\left< \delta_\gamma(\vec{K}-\vec{k'},t')\delta v^*(\vec{k},t'')
 \right>
 \left< \delta
  v(\vec{k'},t')\delta_\gamma^*(\vec{K'}-\vec{k},t'')\right>~.\nonumber \\
\end{eqnarray}
Since the evolution equations for linear order perturbation variables
are independent of $\khat$, we may write
\begin{eqnarray}
 \delta_\gamma(\vec{k},t)&=&\psi_i(\vec{k})\delta_\gamma(k,t)~, \\
 \delta v(\vec{k},t)&=&\psi_i(\vec{k})\delta v(k,t)~, \\
 \Pi (\vec{k},t)&=&\psi_i(\vec{k})\Pi(k,t)~,
\end{eqnarray}
where $\psi_i(\vec{k})$ is the initial perturbation and
$\delta_\gamma(k,t), \delta u(k,t)$, and $\Pi(k,t)$ are the transfer
functions, which are the solution of the Einstein-Boltzmann equations under the
adiabatic initial condition with $\psi_i(\vec{k})=1$. 
By virtue of linearized equations, the overall amplitude and 
spectrum of the primordial fluctuations, $P(k)$, can be given
independently of the transfer functions. Here $P(k)$ is defined
using the two-point correlation function of $\psi_i$ in Fourier space,
\begin{equation}
 \left<\psi_i(\vec{k})\psi_i^*(\vec{k'})\right>=P(k)
 \delta(\vec{k}-\vec{k'})~,
\end{equation}
where $\delta$ is the Dirac delta function. According to the convention,
we write $P(k)\propto k^{n_s-4}$ with $n_s$ being the spectral index of
primordial density perturbations. Using these definitions, we finally obtain, 
\begin{eqnarray}
S(K) &=&
  \left(\frac{4\sigma_T}{3e}\right)^2 
  \int \frac{d^3\vec{k}}{(2\pi)^3}
  K^2 \left[1-(\hat{k}\cdot\hat{K})^2\right] P(|\vec{K}-\vec{k}|) P(k) \nonumber\\
&&\times \biggl\{S_1^2(|\vec{K}-\vec{k}|,k)
-\frac{k}{|\vec{K}-\vec{k}|} S_1(k,|\vec{K}-\vec{k}|)
S_1(|\vec{K}-\vec{k}|,k) \nonumber \\
&&~~~+\frac{1}{4}\left[
	     \beta S_2(k,|\vec{K}-\vec{k}|)S_1(|\vec{K}-\vec{k}|,k)
- \frac{k}{|\vec{K}-\vec{k}|}
\alpha S_1(k,|\vec{K}-\vec{k}|)S_2(|\vec{K}-\vec{k}|,k)\right]
\nonumber \\
&&~~~+\frac{1}{16}\alpha\left[
\alpha S_2^2(|\vec{K}-\vec{k}|,k)-\beta S_2(k,|\vec{K}-\vec{k}|)S_2(|\vec{K}-\vec{k}|,k)
\right]\biggr\}~,
\label{Eq:final}
\end{eqnarray}
where
\begin{eqnarray}
S_1(k,k^\prime)
&=& \int a(t^\prime) \rho_\gamma(t^\prime) \delta_\gamma(k,t^\prime)
         \delta v(k^\prime,t^\prime) dt^\prime ~,\\
S_2(k,k^\prime)
&=& \int a(t^\prime) \rho_\gamma(t^\prime) v_e(k,t^\prime) \Pi_\gamma(k^\prime,t^\prime) dt^\prime ~,\\
\alpha &=& \frac{3K(\hat{K}\cdot\khat)-2k}{3|\vec{K}-\vec{k}|}~,\\
\beta &=& \frac{K\left\{K-k(\hat{K}\cdot\khat)\right\}}{|\vec{K}-\vec{k}|^2}-\frac{2}{3}~.
\end{eqnarray}
Here $S_1$ and $S_2$ describe effects of baryon-photon slip and photon
anisotropic pressure. Specifically, the terms in the second and
fourth lines in Eq.(\ref{Eq:final}) are what we will call in subsequent
sections the contributions to the generation of magnetic fields from
baryon-photon slip and anisotropic stress of photons, respectively.

\subsection{Cosmological Perturbation Theory and Numerical Calculations}
\begin{figure}
\begin{minipage}[m]{0.45\textwidth}
\includegraphics[width=1.0\textwidth]{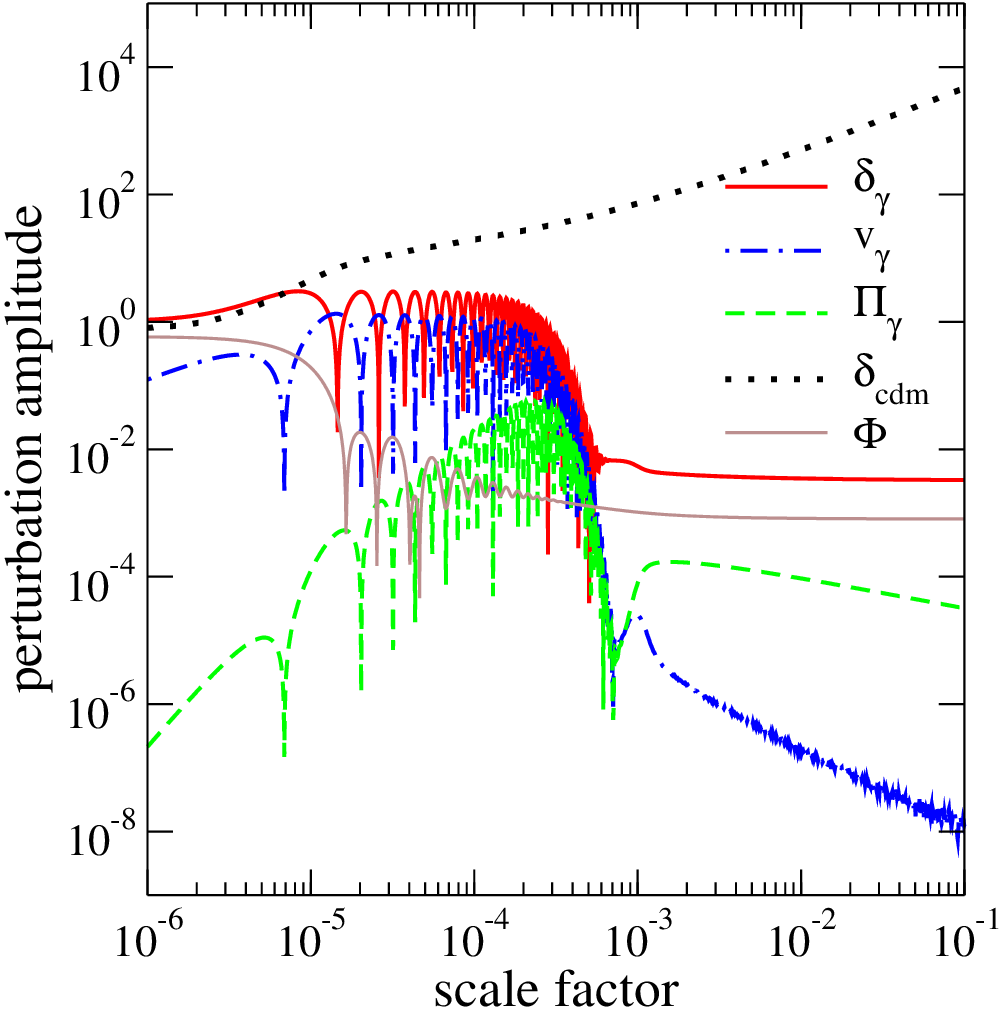}
\end{minipage}
\hspace*{0.02\textwidth}
\begin{minipage}[m]{0.45\textwidth}
\includegraphics[width=1.0\textwidth]{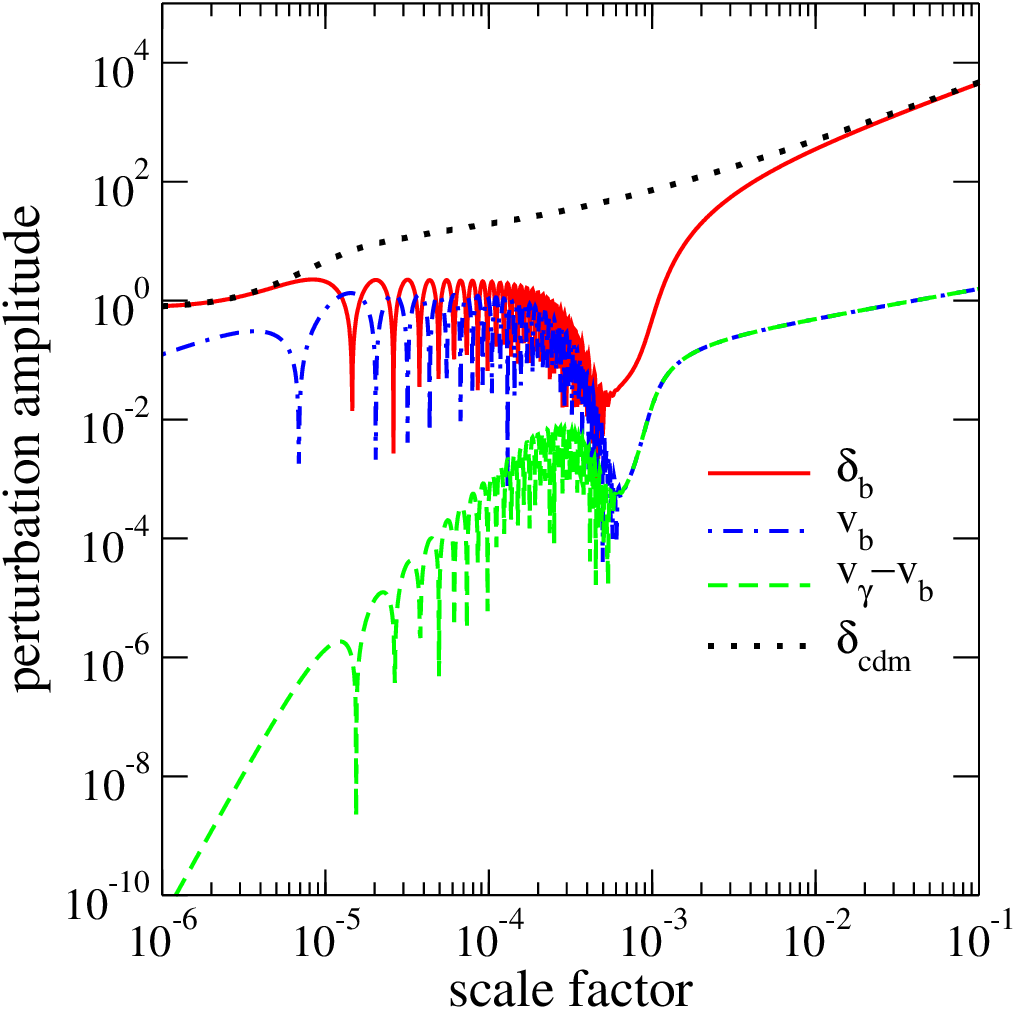}
\end{minipage}
\vspace*{0.2cm}
   \caption{Evolution of perturbations relevant with magnetic field
 generation in a $\Lambda$CDM model with wavenumber $k/h=1.6$
 Mpc$^{-1}$. Left: the five lines represent density contrast of CDM
 ($\delta_{\rm CDM}$; black dotted line), and density contrast
 ($\delta_\gamma$; red line) , velocity perturbation ($v_\gamma$; blue
 dash-dotted line), anisotropic stress ($\Pi_\gamma$; green dashed
 line) of photons, and curvature perturbation ($\Phi$; brown thin line)
 as indicated. Right: time evolutions of density 
 contrast ($\delta_b$; red line) and velocity perturbation ($v_b$; blue
 dash-dotted line) of baryons. 
 The baryon-photon slip ($v_\gamma-v_e$; blue dash-dotted
 line) is also shown in the figure. In the deep radiation dominated era,
 baryon-photon slip and anisotropic stress of photons are severely
 suppressed compared to the other perturbation variables.
 Perturbation variables are shown in
 the Newtonian gauge \cite{1995ApJ...455....7M} while anisotropic stress
 and slip terms are gauge invariant.}
   \label{fig:4}
\vspace*{0.5cm}
\end{figure}

Evolutions of perturbations relevant to the generation of magnetic
fields can be solved within the frame work of the standard
cosmological perturbation theory.  The standard theory describing
evolutions of linear perturbations, which was originally developed by
Lifshiz \cite{Lifshitz:1945du} and summarized in his textbook
\cite{1971ctf..book.....L} and other recent
articles~\cite{1984PThPS..78....1K,1992PhR...215..203M}, is well
tested by a number of observations and firmly established.  As for an
application to cosmological models, the first realistic numerical
calculation was done by Peebles \& Yu \cite{1970ApJ...162..815P},
where the distribution of photons was solved by directory integrating
the Boltzmann equation. The calculations were subsequently extended to
include neutrinos \cite{1983ApJ...274..443B}, and non-baryonic dark
matter \cite{1984ApJ...285L..45B}. These calculations, especially the
resulting angular spectrum of CMB photons, are fully understood within
an analytic treatment by Hu \& Sugiyama \cite{1995ApJ...444..489H}.

\begin{figure}
\begin{minipage}{0.48\textwidth}
\rotatebox{0}{\includegraphics[width=1.0\textwidth]{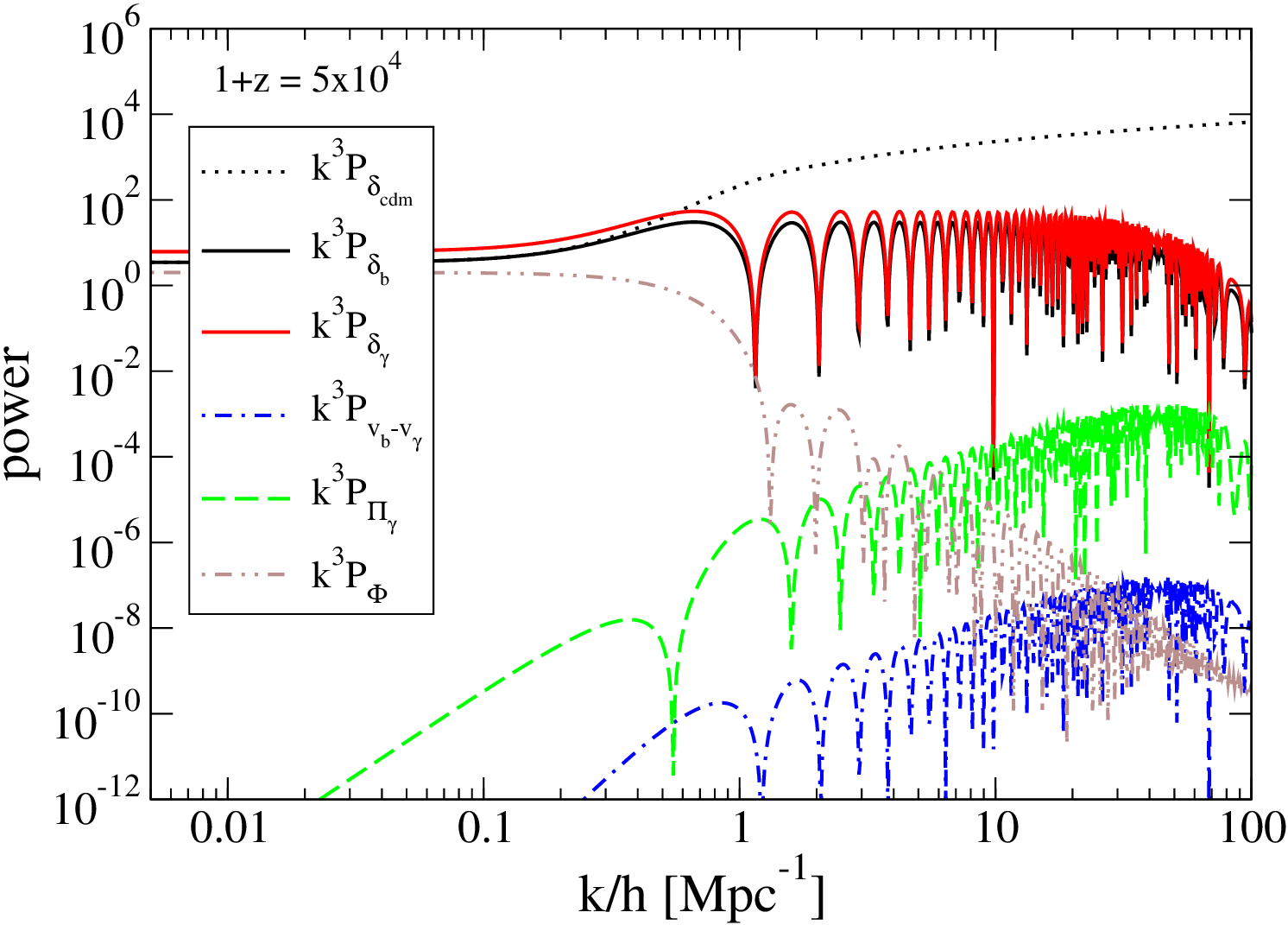}}
\end{minipage}
\hspace*{0.02\textwidth}
\begin{minipage}{0.48\textwidth}
 \rotatebox{0}{\includegraphics[width=1.0\textwidth]{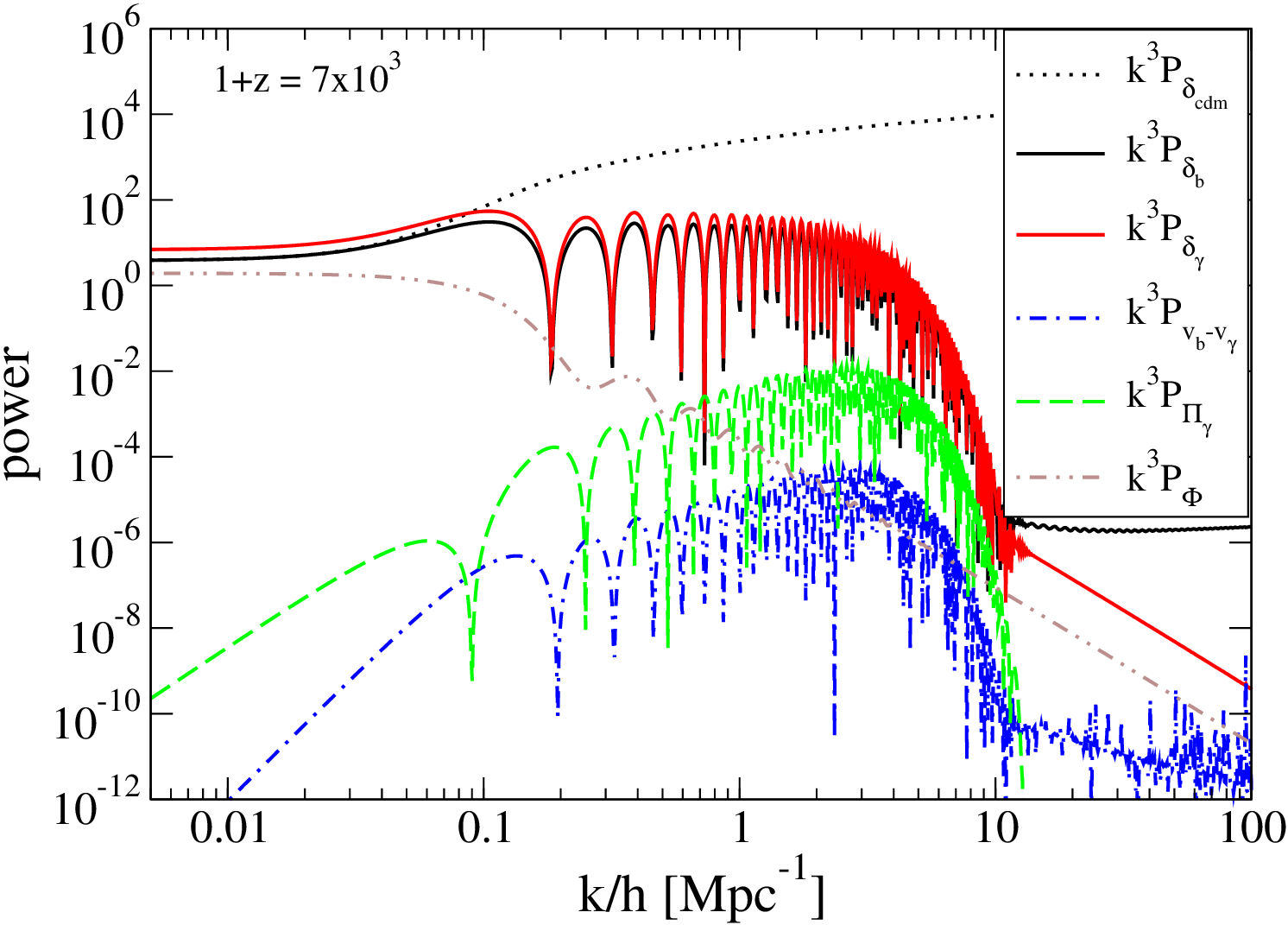}}
\end{minipage}
\begin{minipage}{0.48\textwidth}
\vspace*{0.04\textheight}
\rotatebox{0}{\includegraphics[width=1.0\textwidth]{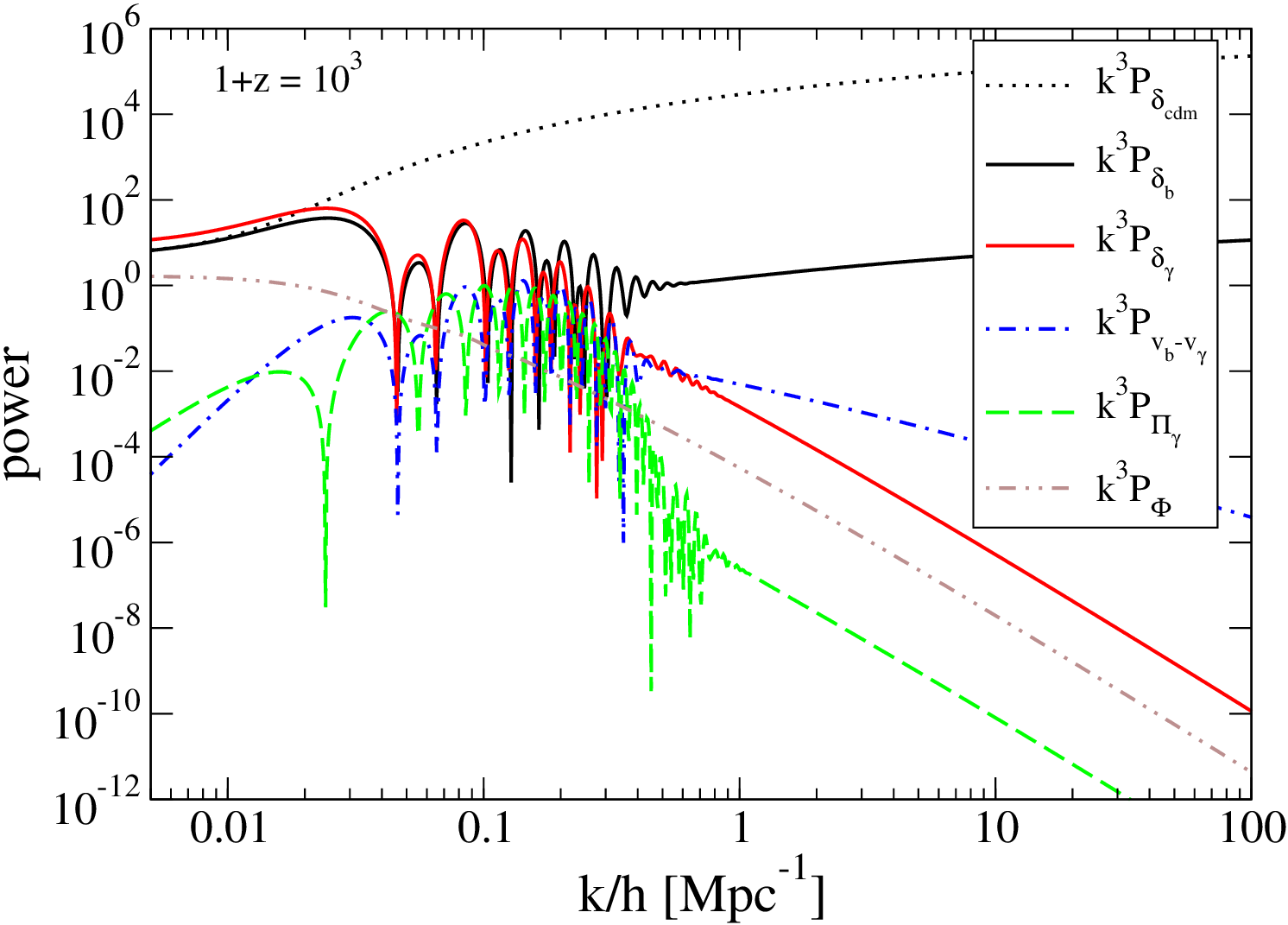}}
\end{minipage}
\hspace*{0.02\textwidth}
\begin{minipage}{0.48\textwidth}
~
\end{minipage}
\vspace*{0.2cm}
\Black{\caption{Spectra of $\delta_{\rm CDM}$ (black dotted line), $\delta_b$
 (thick black line), $\delta_\gamma$ (thin red line), $\Pi_\gamma$
 (green dashed line), $v_b-v_\gamma$ (blue dash-dotted line), and $\Phi$
 (brown dash-dot-dotted line) at redshift
 $z=10^3$ (bottom left), $7\times 10^{3}$ (top right), $5\times 10^{4}$ (top
 left). In general, the spectra at each redshift are divided into four
 parts from large scales to small ones: primordial spectra at super-horizon scales, acoustic oscillation spectra at sub-horizon scales, exponentially damping
 spectra at diffusion scales, and power law phase after the diffusion
 damping. In acoustic oscillation scales at each redshift, the
 amplitudes of density
 perturbations of baryons ($\delta_b$) and photons ($\delta_\gamma$)
 stay constant, while baryon-photon slip ($v_e-v_\gamma$) and
 anisotropic stress of photons ($\Pi_\gamma$) show larger power at relatively
 smaller scales. Powers in baryon-photon slip and anisotropic stress of
 photons get larger at lower redshift.}}
\label{fig:perturbation_spectra}
\vspace*{0.5cm}
\end{figure}

Evolutions of the linear quantities such as $\delta$, $v$, and $\Pi$
in an expanding universe can now be easily solved by the publicly
available Einstein-Boltzmann code such as CMBFAST
\cite{1996ApJ...469..437S} or CAMB \cite{Lewis:1999bs}. In
Fig. \ref{fig:4}, we show an example of the evolution of
perturbations with wavenumber $k/h=1.6$ Mpc$^{-1}$ in standard
$\Lambda$CDM cosmology. Time evolutions can be typically divided into
four characteristic eras.  First, the modes of perturbations are
beyond the cosmic horizon (era I; $a \la 10^{-5}$ in
Fig. \ref{fig:4}). After the horizon crossing, the baryon-photon
fluid undergoes acoustic oscillations (era II; $10^{-5} \la a \la
10^{-3.5}$) until the diffusion of photons erases the perturbations
(era III; $10^{-3.5} \la a \la 10^{-3}$) as the universe expands,
while density perturbations of CDM grow mildly by their own
gravity. After recombination (era IV; $a \ga 10^{-3}$), photons and
baryons are decoupled each other.  In that era photons stream freely
and baryons can evolve according to the gravitational potential of
CDM.

All modes of perturbations in baryons and photons whose wave number is
larger than $\sim 0.1$ Mpc$^{-1}$ are destined to be erased before
recombination by Silk damping~\cite{1968ApJ...151..459S}. Therefore,
each perturbation mode with such wavenumber can contribute to the
generation of magnetic fields only when the mode undergoes acoustic
oscillations. During that era, density and velocity perturbations of
baryons and photons show an oscillatory behavior and their amplitudes
remain almost constant.  Meanwhile the variables suppressed by tight
coupling between photons and electrons, such as $v_e-v_\gamma$ or
$\Pi_\gamma$, grow as the universe expands and the number density of
electrons becomes smaller (see Fig. \ref{fig:4}). Thus the spectra
of perturbations have a peak at the scale of acoustic oscillations or
the scale slightly larger than the diffusion scale at each epoch, that
is clearly shown in Fig. \ref{fig:perturbation_spectra}.

\begin{figure}
\begin{minipage}{0.48\textwidth}
\rotatebox{0}{\includegraphics[width=1.0\textwidth]{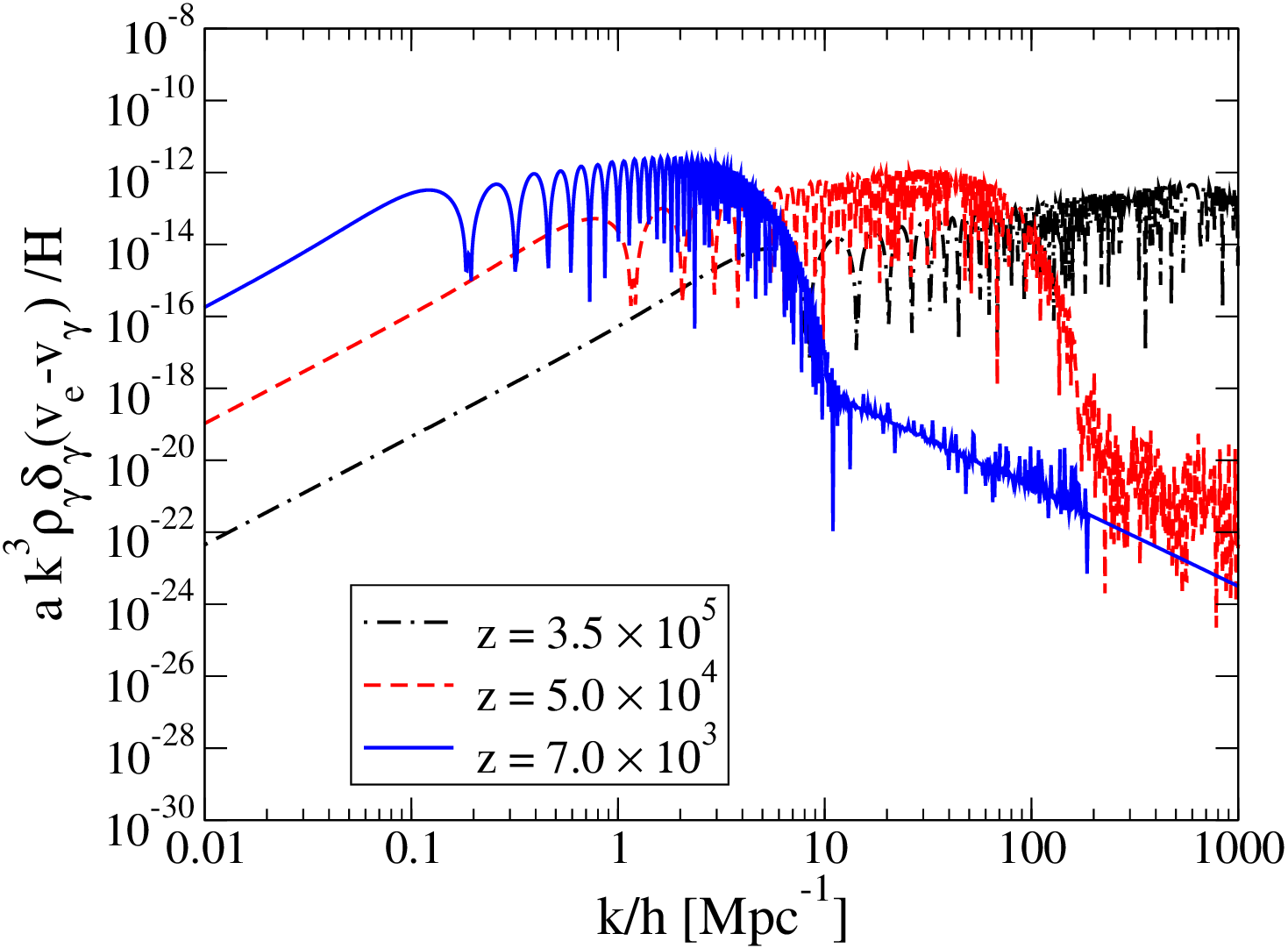}}
\end{minipage}
\hspace*{0.02\textwidth}
\begin{minipage}{0.48\textwidth}
\rotatebox{0}{\includegraphics[width=1.0\textwidth]{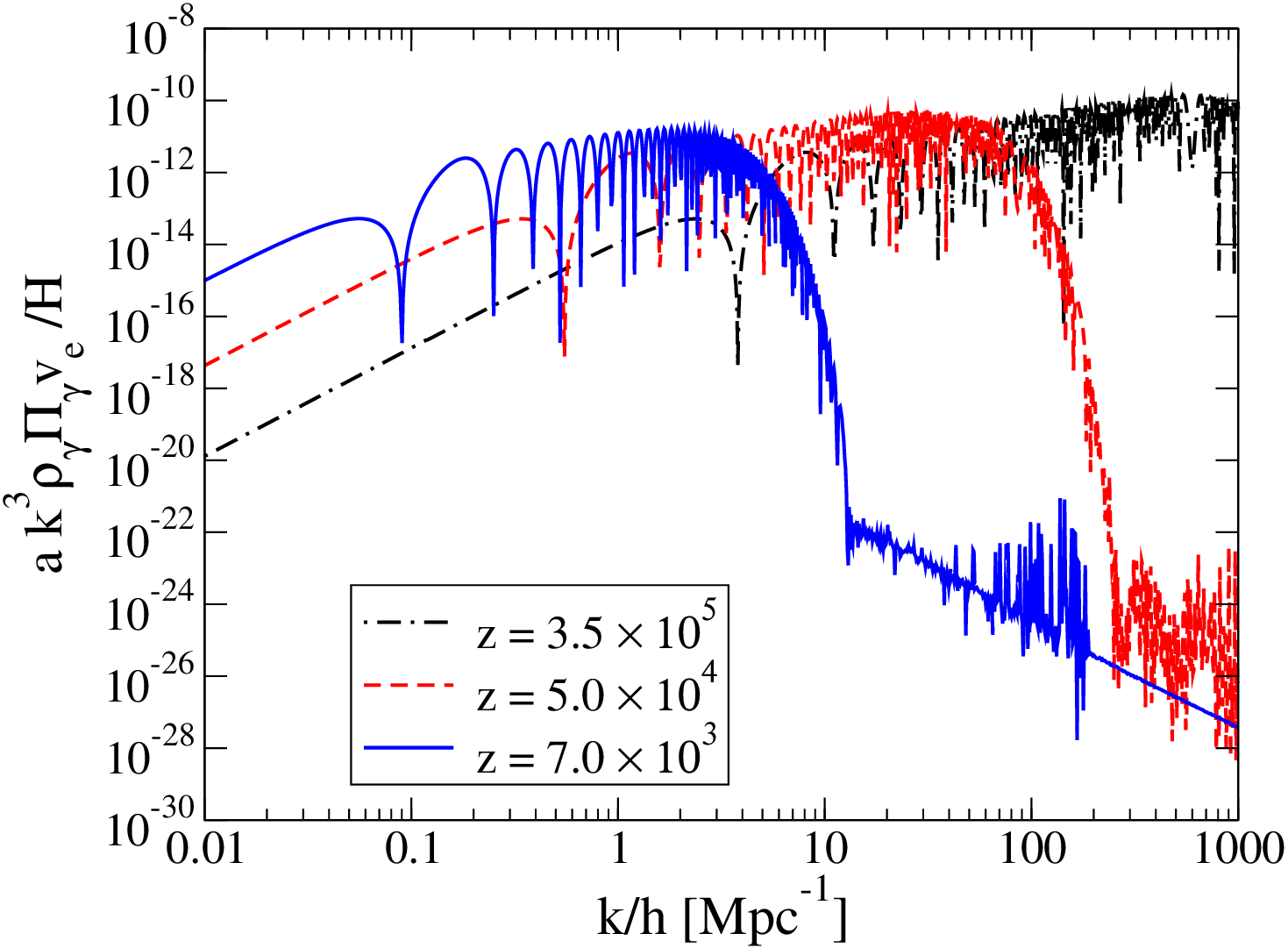}}
\end{minipage}
\caption{Plots of integrand per d$\log(1+z)$,
 $k^3\frac{\rho_\gamma}{H}\delta_\gamma (u_e-u_\gamma)$ (left) and
 $k^3\frac{\rho_\gamma}{H}\Pi_\gamma u_e$ (right). The magnetic field
 spectrum results from non-linear convolution and time integration of
 these spectra. The contributions mainly comes from the waves between
 the scales of cosmic horizon, and the Silk diffusion at each redshift.} 
\label{fig:source_terms}
\vspace*{0.5cm}
\end{figure}

We plot $\left(a\rho_\gamma/H\right) k^3 P(k) \delta v(k)
\delta_\gamma(k)$ and $\left(a\rho_\gamma/H\right)k^3 P(k) v_b(k)
\Pi_\gamma(k)$ in the left and right panels of
Fig.~\ref{fig:source_terms}. These spectra naively correspond to the
source terms $S_1$ and $S_2$ in Eq. (\ref{Eq:final}). 
Of course, the magnetic field spectrum should be obtained by a
non-linear convolution of these spectra given by Eq. (\ref{Eq:final}).
However, because the smaller scale modes than the diffusion scale are
absent at each redshift and thus the non-linear couplings mainly come
from the modes with comparable wavenumbers, one can expect that the
cosmologically generated magnetic fields will roughly have a spectrum
similar to the envelope curves of these spectra.

\begin{figure}
\begin{minipage}[m]{0.48\textwidth}
\includegraphics[width=1.0\textwidth]{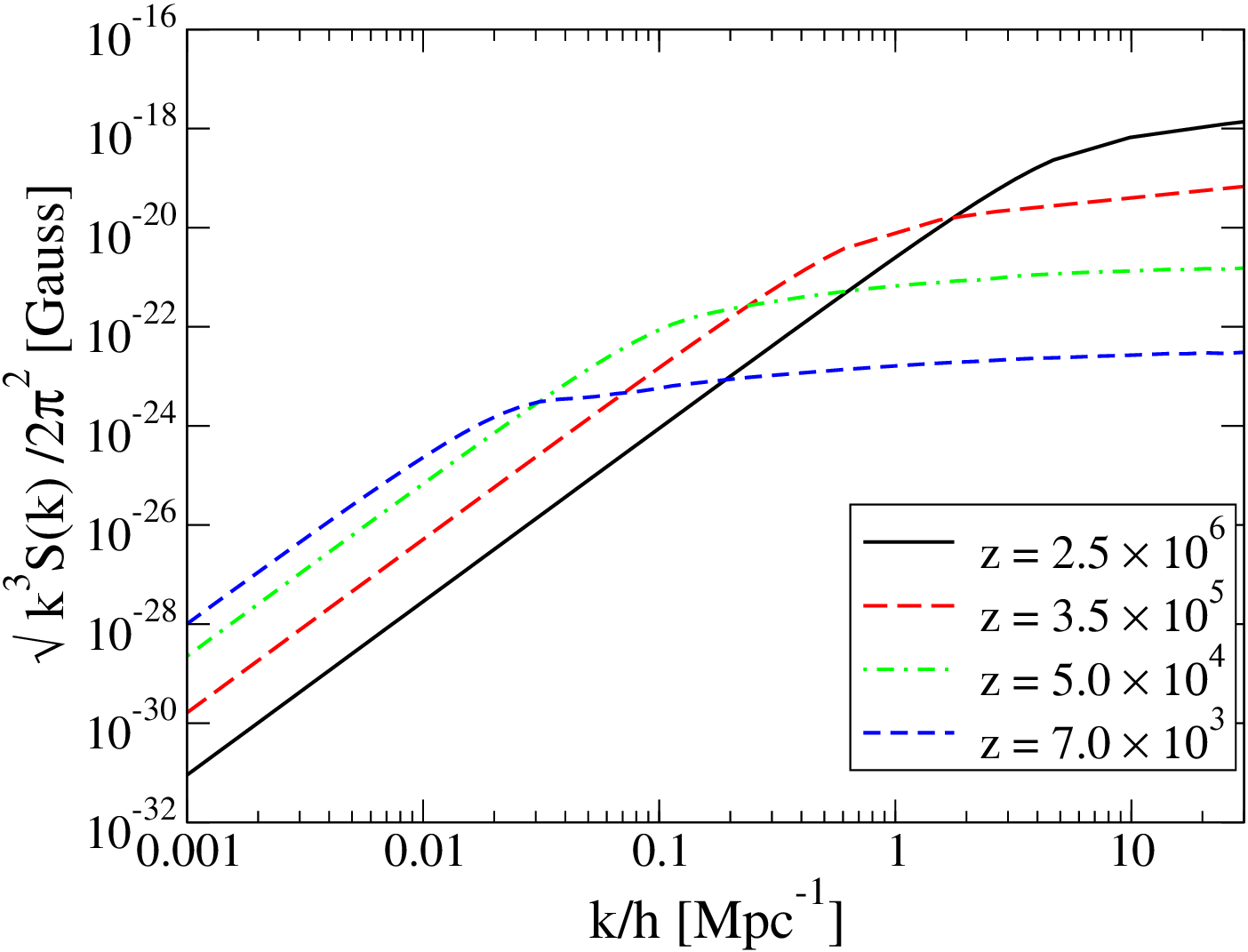}
\end{minipage}
\hspace*{0.02\textwidth}
\begin{minipage}[m]{0.48\textwidth}
\includegraphics[width=1.0\textwidth]{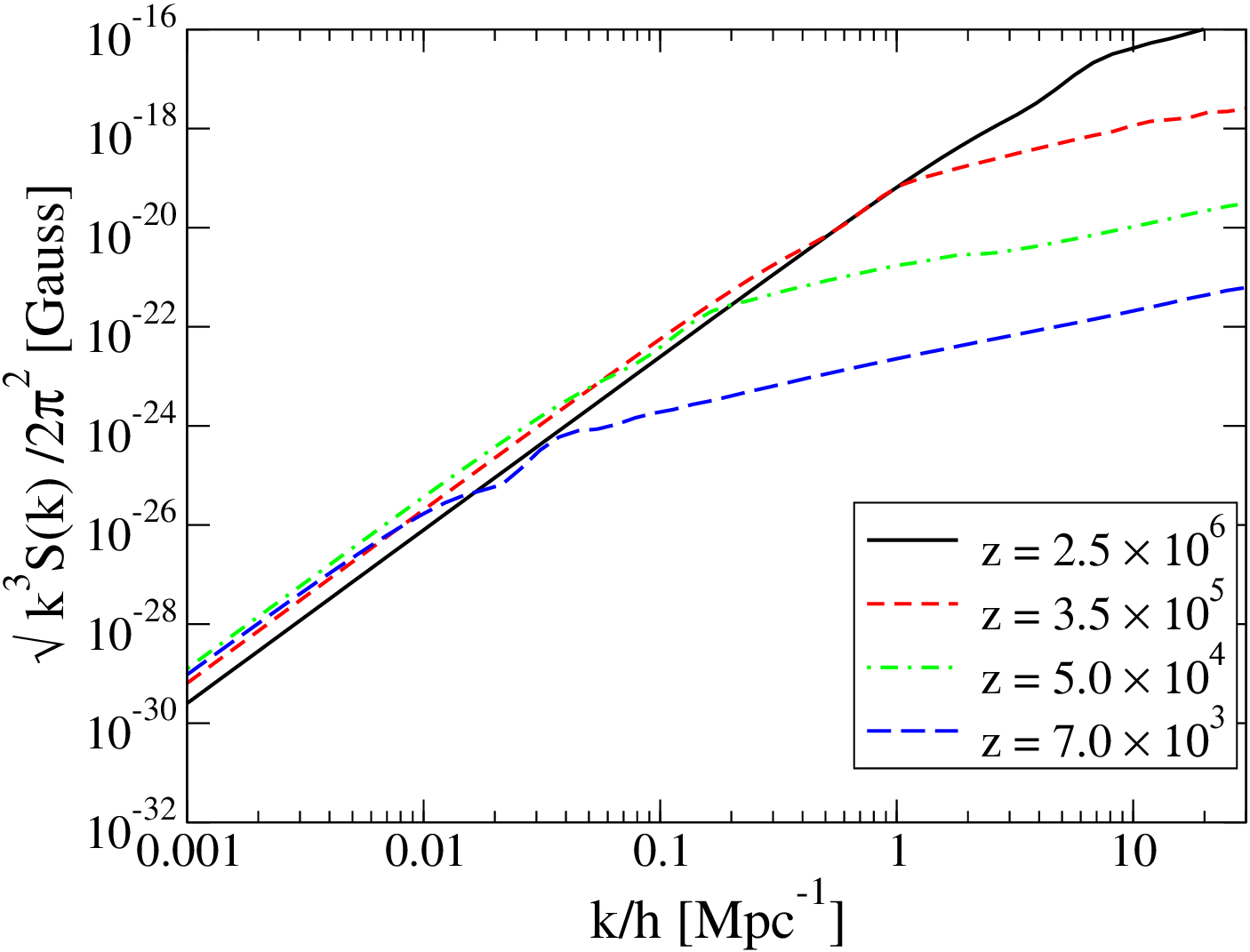}
\end{minipage}
   \caption{Time evolution of Spectra of conformal magnetic fields
 generated from baryon-photon slip (left) and anisotropic stress of
 photons (right). Corresponding redshifts are also indicated in the
 figures. For both contributions, the smaller scale fields are created
 at the higher redshift.}\label{fig:2}
\vspace*{0.5cm}
\end{figure}

\begin{figure}
\begin{minipage}[m]{0.48\textwidth}
\includegraphics[width=1.0\textwidth]{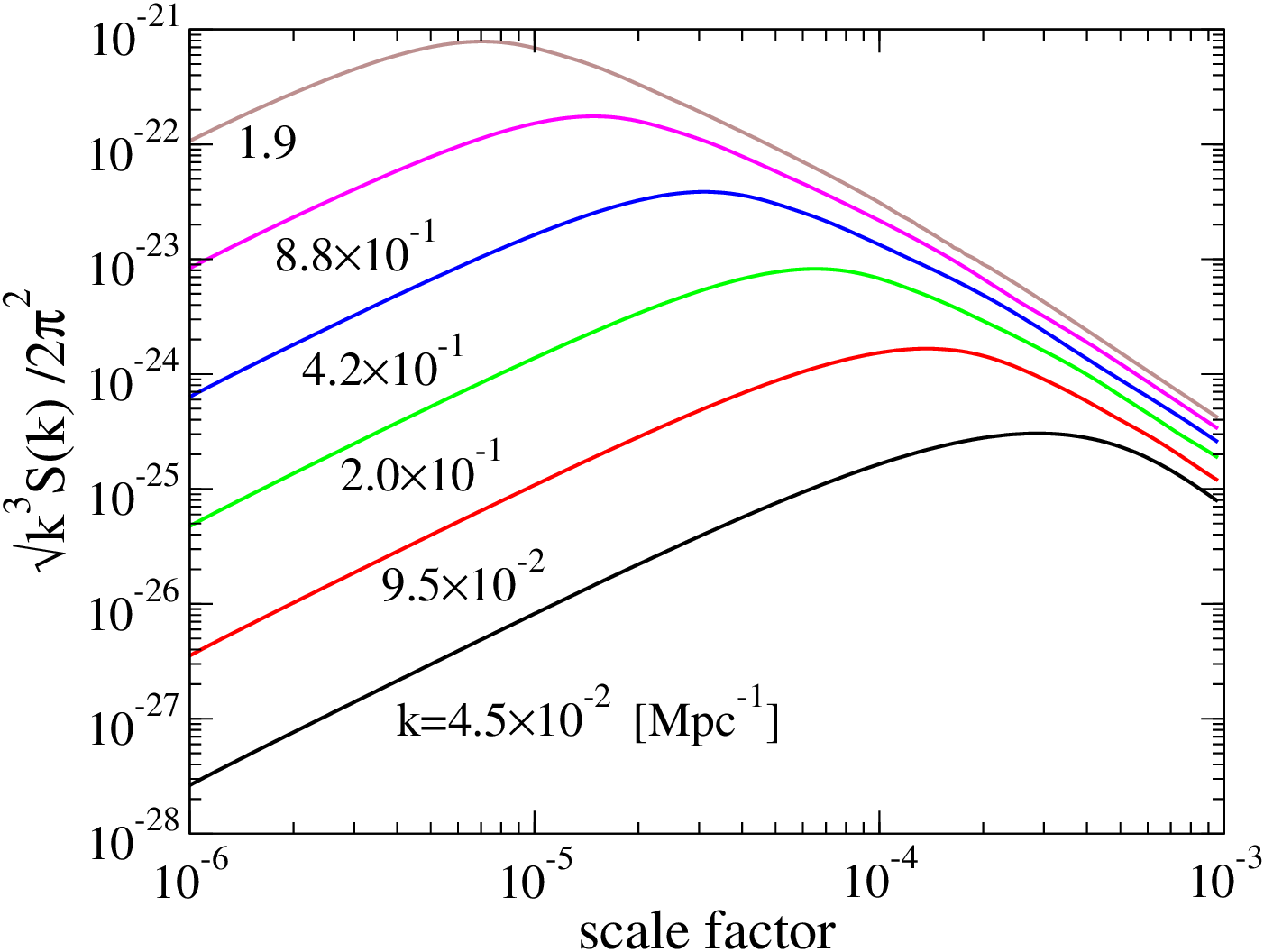}
\end{minipage}
\hspace*{0.02\textwidth}
\begin{minipage}[m]{0.48\textwidth}
\includegraphics[width=1.0\textwidth]{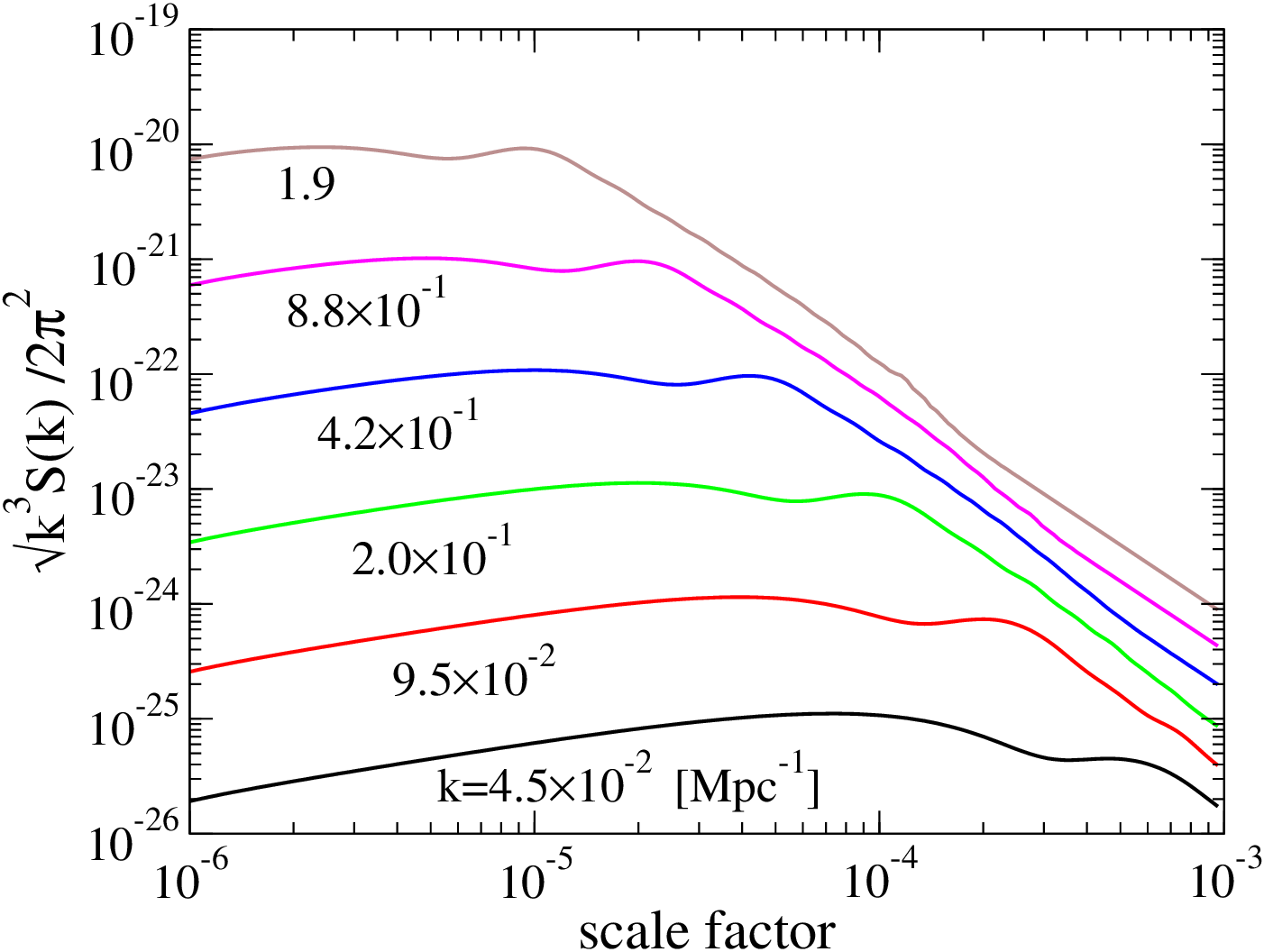}
\end{minipage}
   \caption{Time evolution of magnetic fields in units of Gauss,
 $\sqrt{k^3 S(k)/2\pi^2}$, generated from baryon-photon slip (left) and
 anisotropic stress of  photons (right) with fixed wavenumbers as
 indicated. Magnetic fields at smaller  scales have been created earlier
 and their amplitudes are larger. After  their generation magnetic
 fields diminish as $a^{-2}$ adiabatically. 
}
   \label{fig:3}
\end{figure}

We numerically integrated Eq.({\ref{Eq:final}}) to obtain the magnetic
field spectra. The cosmological parameters in our
calculation are fixed to the standard $\Lambda$ CDM values in a flat
universe, i.e., 
\begin{equation}
(h,n_s,\Omega_b,\Omega_\Lambda,A)=(0.7,~1.0,~0.04,~0.70,~2.7\times 10^{-9})~,
\end{equation}
where $h$ is the current Hubble parameter in units of $100$km/s/Mpc,
$n_s$ is the power spectrum index of primordial density fluctuations,
$\Omega_b$ and $\Omega_\Lambda$ are energy densities in baryon and
cosmological constant in units of the critical density, and $A$ is
overall amplitude of density perturbation squared.  Magnetic field
spectra induced by baryon-photon slip and anisotropic stress of
photons at different redshifts, which correspond to second and fourth
lines of Eq.(\ref{Eq:final}), respectively, are shown in left and
right panels of Fig. \ref{fig:2}.  In these figures we take the
combination of $\sqrt{k^3 \left<BB^*\right>}/2\pi^2$ to show in units
of Gauss. The spectra consist of two slopes. At super-horizon scales
the slope is proportional to $k^{3.5}$ for both contributions while it
becomes less steep at sub-horizon scales. The turning points of the
spectra roughly correspond to the Hubble horizon scale at each
redshift while it may be difficult to see clear turnoff in the spectra
generated from anisotropic stress (right panel).  
Clearly, one can find that small scale magnetic fields are
created at higher redshifts. After their generation, magnetic fields
$B$ are adiabatically diminishing as $a^{-2}$.  To see this fact more
explicitly, we depict the time evolution of magnetic field with fixed
wavenumber $K$ in Fig. \ref{fig:3}. Analytic interpretation for
these magnetic field spectra is given in the next section.

\section{Analytic Interpretation of the Magnetic Field Spectrum}
\subsection{Magnetic Field Spectrum at Large Scales}
Magnetic fields are mainly created after the modes of perturbations
with the corresponding scale enter the cosmic horizon and are causally
contacted.  Magnetic fields at larger scales than the cosmic horizon
are generated only as a consequence of non-linear couplings of each
Fourier mode.  In the previous studies it was suggested that at
largest scales the magnetic field spectrum generated from density
perturbations has the power proportional to $k^2$
\cite{2005PhRvD..71d3502M}. If one carefully evaluates, however, the
terms proportional to $k^2$ exactly cancel each other out, and the
spectrum starts from the term proportional to $k^4$
as we shall show below. 
In order to see this, let us consider the spectrum generated by the first term
(baryon-photon slip term) in equation (\ref{eq:B_dot}),
\begin{eqnarray}
 \left<B^i(\vec{K},t)B_i^*(\vec{K'},t)\right>&\propto&
  \int d^3\vec{k}  
    K^2 \left\{1-(\hat{k}\cdot\hat{K})^2\right\} P(|\vec{K}-\vec{k}|)
    P(k) \nonumber \\
    && \times \biggl\{S_1^2(|\vec{K}-\vec{k}|,k)
-\frac{k}{|\vec{K}-\vec{k}|} S_1(k,|\vec{K}-\vec{k}|)
S_1(|\vec{K}-\vec{k}|,k) \biggr\}~.
\label{Eq:(67)}
\end{eqnarray}
In the limit $K/k \to 0$, we can write
\begin{eqnarray}
 |\vec{K}-\vec{k}|&=&\sqrt{k^2-2kK\gamma+K^2} \approx
  k-K\gamma~, \nonumber \\
 |\vec{K}-\vec{k}|^{-1}&\approx&k^{-1}\left(1+\frac{K}{k}\gamma\right)~, \nonumber
\end{eqnarray}
where $\gamma \equiv \hat{k}\cdot\hat{K}$. Furthermore, time
integration terms can be treated as $K$ independent in the same limit,
i.e., $S_1(|\vec{K}-\vec{k}|,k) \approx S_1(k,|\vec{K}-\vec{k}|)
\approx T(k)$. Remembering that the spectrum of density perturbations is
written as $P(k)\propto k^{n_s -4}$, we can rewrite equation (\ref{Eq:(67)}) as
\begin{eqnarray}
\left<B^i(\vec{K},t) B_i^*(\vec{K'},t)\right>
&\propto& \int k^2 dk \int^{1}_{-1} d\gamma (1-\gamma^2) P(|\vec{K}-\vec{k}|)P(k)
          K^2 \left(1-\frac{k}{|\vec{K}-\vec{k}|}\right) T^2(k) \nonumber\\
&\approx& \int k^2dk \int^{1}_{-1} d\gamma (\gamma-\gamma^3)
          \left[k-(n_s - 4) K \gamma\right] k^{n_s - 4}
          \left(-\frac{K^3}{k}\right) T^2(k) \nonumber \\
&\propto& K^4~.
\end{eqnarray}
We can see this non-linear power law tail at large scales in Fig. \ref{fig:2}.

\subsection{Magnetic Field Spectrum at Sub-horizon Scales}
In this subsection we show that the behavior of the magnetic field
spectrum at small scales can be understood within the standard theory
of cosmological perturbations.  Since Compton scattering is an
essential process for generation of magnetic fields, magnetic fields
start to be generated when the relevant modes come across the horizon
and are causally connected. On the other hand, once the modes become
shorter than the diffusion scale of photons, magnetic fields 
cannot be no longer generated.  Therefore 
magnetic fields with wavenumber $K$ are mainly
created from density perturbations with wavenumber $k \sim K \la
k_{\rm diff}$, where $k_{\rm diff}$ is the corresponding wavenumber of
the diffusion scale at each time.  Within the approximation that
$k\sim K$, $|\vec{K}-\vec{k}|\sim K$, the spectrum of magnetic fields is
roughly 
given by
\begin{equation}
 S(k) \sim \left(\frac{4\sigma_T}{3e}\right)^2 \left(\int d^3k k^2 s_1^2
					       +\int d^3k k^2 s_2^2\right)~,
\end{equation}
where
\begin{eqnarray}
s_1&\equiv& P(k)S_1(k,k)~,\\
s_2&\equiv& P(k)S_2(k,k)~.
\end{eqnarray}
Here we neglected sub-dominant cross correlating terms.
We have already plotted $k^3 s_1 \propto k^3 \frac{a\rho_\gamma}{H} P(k) \delta v(k)
\delta_\gamma$ and $k^3 s_2 \propto k^3 \frac{a\rho_\gamma}{H} P(k) v_b(k)
\Pi_\gamma$ in the left and right panels of Fig. \ref{fig:source_terms}.
These spectra indeed indicate that magnetic fields are created when the modes
of perturbations come across the cosmic horizon and undergo acoustic
oscillations. 
In fact, the largest contributions come from the modes of
perturbations slightly larger than the diffusion scale.

The behavior of acoustic oscillations in the early
universe can be analytically understood by using tight coupling
approximations. In the tight coupling regime, Compton scattering occurs
sufficient enough so that photons and baryons (electrons) can behave as
a coupled fluid. Therefore, the baryon-photon slip ($v_b -
v_\gamma$) and the anisotropic stress of photons
$\Pi_\gamma$ are severely suppressed. Specifically, within the tight
coupling approximation \cite{1995ApJ...444..489H} the perturbation variables
satisfy, 
\begin{eqnarray}
v_b &\approx& v_\gamma \approx \delta_\gamma~, \label{eq:acoustic_approx}\\
v_b - v_\gamma &\approx& \epsilon v_\gamma
 \left(\frac{3\rho_b}{4\rho_\gamma}\right)\approx \epsilon \delta_\gamma
 \left(\frac{3\rho_b}{4\rho_\gamma}\right)~,\label{eq:vb-vg_approx}\\
\Pi_\gamma &\approx& \epsilon v_\gamma \approx \epsilon \delta_\gamma~, \label{eq:pi_g_approx}
\end{eqnarray}
where $\epsilon$ is a tight coupling parameter roughly given as
\begin{equation}
\epsilon = \frac{k}{a n_e \sigma_T} \sim 10^{-2}\left(\frac{k}{1{\rm Mpc}^{-1}}\right)\left(\frac{1+z}{10^4}\right)^{-2}\left(\frac{\Omega_b
	 h^2}{0.04}\right)^{-1}~.
\label{eq:tight_coupling_param}
\end{equation}
This parameter represents how tightly photons are coupled with
electrons and gives a criteria whether one can treat photons and
electrons as a single perfect fluid. So, the parameter in equations
(\ref{eq:vb-vg_approx}) and (\ref{eq:pi_g_approx}) simply reflects the
fact that when the photons and electrons can be treated as a tightly
coupled single fluid (i.e., when the tight coupling parameter
$\epsilon$ is sufficiently small), the baryon-photon slip $(v_e -
v_\gamma)$, and the anisotropy of the distribution function
($\Pi_\gamma$) are severely suppressed in comparison with the density
fluctuation ($\delta_{\gamma}$) (see Fig.~\ref{fig:4}).

Let us see how the spectra in Fig.~\ref{fig:source_terms} can be
understood using the above relations. Noting that Hubble parameter
scales as $H \propto a^{-2}$ in the radiation dominated era, we have the
following relations 
\begin{eqnarray}
\frac{\rho_\gamma}{H}(v_b - v_\gamma) &\propto&
 \frac{a^{-4}}{a^{-2}}\epsilon v_\gamma
 \left(\frac{3\rho_b}{4\rho_\gamma}\right) \propto k \delta_\gamma
 a(k)~,\\
\frac{\rho_\gamma}{H}\Pi_\gamma &\propto& \frac{a^{-4}}{a^{-2}}
 \epsilon v_\gamma \propto k a^{0} \delta_\gamma~,
\end{eqnarray}
where we have used the tight coupling relations
(\ref{eq:acoustic_approx}), (\ref{eq:vb-vg_approx}), and
(\ref{eq:pi_g_approx}). 
In the case where the spectrum of primordial density perturbations is
scale invariant, $k^3 P(k) \delta \delta$ is constant when the mode
with wave number $k$ enters the Hubble horizon. Therefore, the spectra in
Fig. \ref{fig:source_terms} should be proportional to
\begin{eqnarray}
k^3 s_1 &\propto& k^3 \frac{a \rho_\gamma}{H} P(k) \delta_\gamma
 (v_b(k)-v_\gamma(k)) \propto k a^2~,  \label{eq:k3s1}\\
k^3 s_2 &\propto& k^3 P(k) \Pi_\gamma v_b \frac{a \rho_\gamma}{H} \propto
 k a~. \label{eq:k3s2}
\end{eqnarray}
These behaviors are clearly seen in Fig. \ref{fig:source_terms}. 

However, a complication arises when considering the spectrum of magnetic
fields. Magnetic fields can not be generated from the first order
solution of tight coupling approximation \cite{futurework1}. For example, 
one finds that the contribution from the cross product between the
gradient of density perturbation of photons ($\delta_{\gamma,j}$) and
velocity differences of electrons and photons ($v_{ej}-v_{\gamma j}$) 
 (the first term in Eq.(\ref{eq:B_dot})) vanishes because these
vectors are colinear in the first order tight coupling approximation.
 Therefore, we should consider
the second order solutions of tight coupling expansion which are
proportional to ${\cal O}(\epsilon^2)$. 
Accordingly, the part of the spectra which can contribute to magnetic
fields (Eqs. (\ref{eq:k3s1}) and (\ref{eq:k3s2})) should now be 
evaluated as
\begin{eqnarray}
k^3 s_1|_{\rm mag} &\propto& k^3 \frac{a \rho_\gamma}{H} P(k) \delta_\gamma
 (v_b(k)-v_\gamma(k))|_{\rm mag} \propto \epsilon\times k a^2\propto k^2
 a^4 ~,  \\
k^3 s_2|_{\rm mag} &\propto& k^3 P(k) \Pi_\gamma v_b \frac{a
 \rho_\gamma}{H}|_{\rm mag} \propto \epsilon\times k a \propto k^2 a^3~.
\end{eqnarray}
From these relations we found that the contribution to magnetic
fields becomes larger as the universe expands larger (the larger $a$). 
This means that the scale factor $a$ should be evaluated at 
the onset of the photon diffusion where the density perturbations are
erased, because the contribution becomes largest at that time.
The diffusion length is given by  \cite{1995ApJ...444..489H}
\begin{equation}
k_{\rm
 diff}^{-2}=\frac{1}{6}\int\frac{1}{an_e\sigma_T}\frac{\frac{16}{15}+\frac{R^2}{1+R}}{1+R} \frac{dt}{a}~,
\end{equation}
from which one obtains the relation between the diffusion length and scale
factor in the radiation dominated era:
$k_{\rm diff}= 4.3\times \left(\frac{1+z}{10^4}\right)^{3/2}$ Mpc$^{-1}$
$\propto a^{-3/2}$. Combining these considerations altogether, 
the spectrum of magnetic fields at sub horizon scales is
roughly given by, 
\begin{eqnarray}
S(k) &\sim& k^5 s_1^2 + k^5 s_2^2 \nonumber \\
 &\sim& k^{-7/3} ~(\mbox{slip term})~+ k^{-1}~(\mbox{$\Pi_\gamma$ term})~.
\end{eqnarray}
At smaller scales anisotropic stress of photons ($\Pi_\gamma$) should be
the dominant contributor to the magnetic fields.

\subsection{Cut-off of the magnetic field spectrum}
As we discussed in the previous subsection, the strength of magnetic
fields increases as $B \sim \sqrt{k^3 S(k)} \propto k$ on small scales
where the contribution from anisotropic stress of photons is dominant.
One might expect there exists more and more magnetic fields on smaller
scales.  However careful treatment is needed for the generation of 
magnetic fields on very small scales since 
the relativistic effects of  
electrons, which we omit in the formulation developed in section
II,  play an important role 
when the temperature of the universe was higher than the electron
rest mass, i.e., $T  > 511{\rm keV}$.  The power spectrum of 
magnetic fields below the diffusion scale at  $T = 511{\rm keV}$,
therefore, would be modified. 

There are two potential effects which affect the generation of
magnetic fields quantitatively when the temperature of the universe
was higher than the electron rest mass and electrons were
relativistic. First one is the transition of scatterings between
photons and electrons from the Thomson regime to the Compton one. The
cross section of Compton scattering is proportional to the inverse
square of the center of mass energy while that of Thomson scattering
is independent of the energy. At first glance, this transition would
result in the reduction of the electric current and hence the
reduction of magnetic fields because photons can not push electrons
effectively (Eq. \ref{eq:B_dot}).  Interestingly, however, the weaker
interaction makes the velocity difference between photons and
electrons and the anisotropic stress of photons larger, and these two
effects would cancel each other out.  Specifically baryon-photon
slip and anisotropic stress of photons responsible for the generation
of magnetic fields are proportional to the tight coupling parameter
$\epsilon$ in equation (\ref{eq:tight_coupling_param}), which contains
$\sigma_T$ in the denominator. Inserting this into equation
(\ref{eq:B_dot}), we find that magnetic fields generated from
cosmological perturbations would be cross section independent, i.e., 
\begin{equation}
B \propto \sigma_T^{0}~.
\end{equation}

The second relativistic effect to be considered is the
electron-positron pair creation from photons through $2 \gamma
\leftrightarrow e^{+} + e^{-}$.  In the thermal history of the
standard big bang paradigm, electrons and positrons were in thermal
equilibrium with photons before the temperature of the universe was
around and higher than the electron rest mass.  At that time their
number densities were about the same as that of photons. As the
universe expanded and cooled, photons became less energetic and
unable to pair-create electron-positron pairs while existing
electron-positron pairs annihilated into photons. The number
density of electrons decreased drastically when the universe cooled
below the critical temperature corresponding to the electron rest
mass, and consequently, tiny residual number of electrons compared to
that of photons, $n_e/n_\gamma \sim 10^{-9}$, has survived.  In other
words, if one looks back in the thermal history of the universe, the
number density of electrons increased all of sudden by a factor
of $10^{9}$ around the critical temperature.
This effect would significantly modify the magnetic
field spectrum. Since couplings between photons and electrons (and
positrons) were so tight above the critical temperature than 
below, the relative velocity between them or
anisotropic stress of photons, and thus induced electric current, were
smaller above the critical 
temperature. Specifically, the tight coupling parameter is proportional
to the inverse of electron number density, so are the magnetic fields, 
\begin{equation}
B \propto n_{e}^{-1}~,
\label{eq:SI}
\end{equation}
which leads to the reduction of the amplitude of the magnetic
field spectrum  by a factor of $10^{-9}$ at the corresponding scale.

The diffusion scale at the critical temperature ($T =511\rm keV$) 
is $\sim 2\times 10^{-3}$ pc in the comoving scale.  Therefore we expect that 
the spectrum of magnetic fields continues to increase up to
this scale and rapidly drops on this scale.  There exists
negligible amount of magnetic fields on smaller scales.  
To summarize, the magnetic field spectrum generated from scale-invariant
density perturbations is given as
\begin{equation}
  B \propto \sqrt{k^3 S(k)} \propto \left\{
  \begin{array}{rll}
     &k^{3.5} & \mbox{\hspace*{1cm}(\phantom{$10$ ~Mpc$ \la$ }$\lambda \ga 100$ Mpc)} \\
     &k^{1/3} & \mbox{\hspace*{1cm}($10$ ~Mpc$ \la \lambda \la 100$ Mpc)} \\
     &k & \mbox{\hspace*{1cm}($10^{-3}$~ pc$\la \lambda \la 10~~$ Mpc)} \\
     &\sim  0 & \mbox{\hspace*{1cm}(\phantom{$10$ ~Mpc$ \la$ }$\lambda
      \la 10^{-3}$ pc)}
  \end{array} \right.
\end{equation}
\begin{figure}
\includegraphics[width=0.7\textwidth]{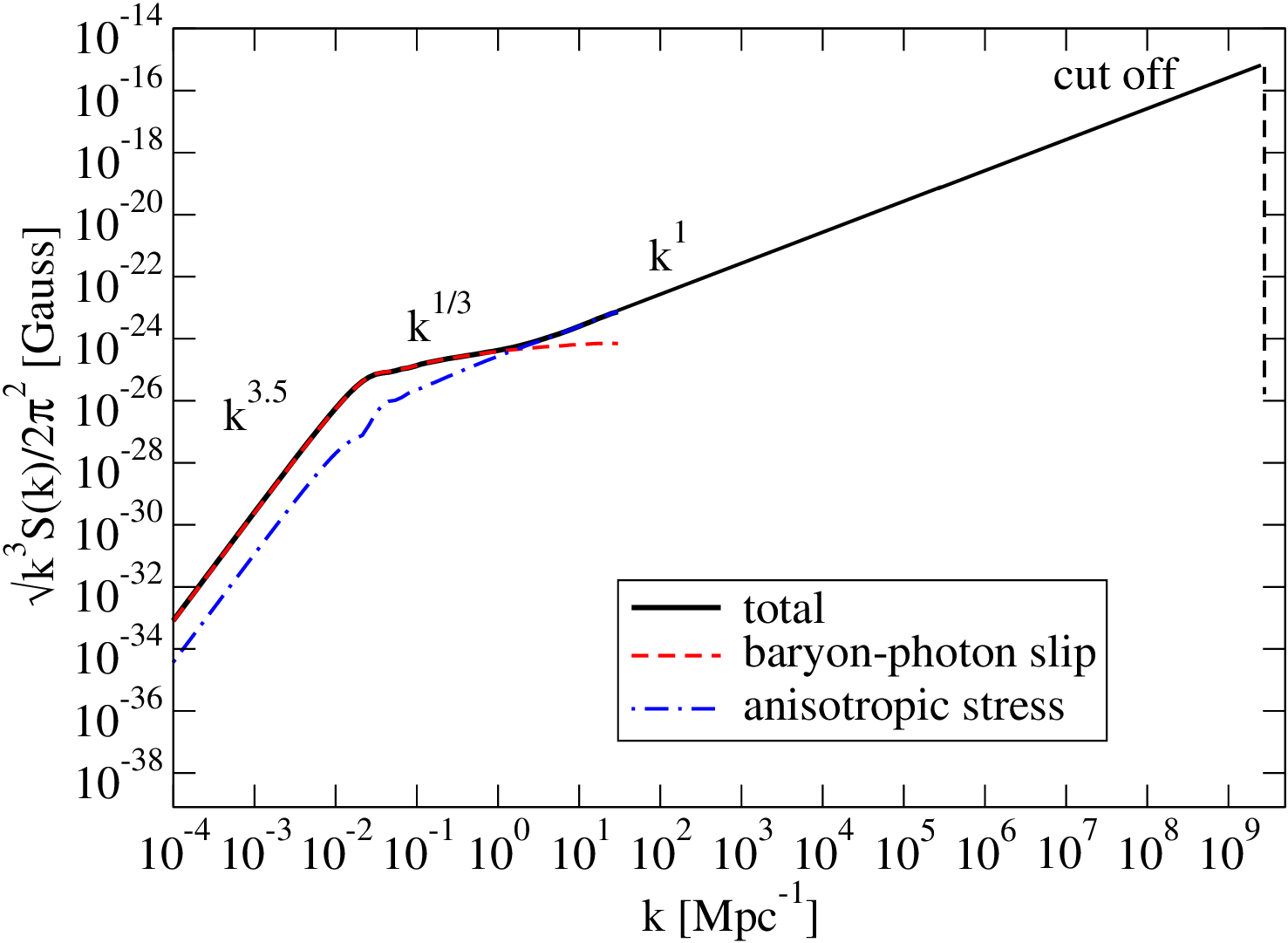}
\caption{Spectrum of magnetic fields generated from cosmological density
 perturbations at $z=1100$. Red dashed line shows contribution
 from baryon photon slip, blue dash-dotted line from anisotropic stress
 of photons, and black like shows the total spectrum. The spectrum was
 obtained by direct numerical integration up to k=30 Mpc$^{-1}$, and by
 analytic extrapolation above it. The spectrum has a slope proportional
 to $k^{3.5}$ at larger scale than the horizon at recombination. At smaller
 scales than $\sim $1 Mpc$^{-1}$, magnetic fields are dominated by the
 contribution from the anisotropic stress of photons.
At smallest scale around $2\times 10^{-3}$ pc, the spectrum would have a
 cut-off due to the relativistic effects of electrons (and positrons).}
\label{fig:6}
\end{figure}
We depict the spectrum of magnetic fields in Fig. \ref{fig:6}. By
remembering the fact that the spectrum of magnetic field $S(k)$ is
proportional to $P(k)P(k-k^\prime)$, the spectrum can be easily
generalized in the case of tilted spectrum:
\begin{equation}
  B \propto \sqrt{k^3 S(k)} \propto \left\{
  \begin{array}{rll}
     &k^{3.5} & \mbox{\hspace*{1cm}(\phantom{$10$ ~Mpc$ \la$ }$\lambda \ga 100$ Mpc)} \\
     &k^{n_s-2/3} & \mbox{\hspace*{1cm}($10$ ~Mpc$ \la \lambda \la 100$ Mpc)} \\
     &k^{n_s} & \mbox{\hspace*{1cm}($10^{-3}$~ pc$\la \lambda \la 10~~$ Mpc)} \\
     &\sim  0 & \mbox{\hspace*{1cm}(\phantom{$10$ ~Mpc$ \la$ }$\lambda \la 10^{-3}$ pc)}
  \end{array} \right.
\end{equation}
where $n_s$ is the spectral index of primordial density perturbations.
\section{Magnetic Helicity}
So far, we have investigated the spectrum of magnetic field amplitude
while there exists another important quantity, helicity, which defines
the property of stochastic magnetic fields.  Magnetic
helicity is a quantity which characterizes magnetic field
configuration how much the magnetic field lines are twisted, or how
many closed magnetic lines are linked \cite{1993noma.book.....B}.
Among the cosmological mechanisms to generate magnetic fields, a
number of scenarios predicts the primordial fields with non-zero
helicity
\cite{2001PhRvL..87y1302V,2000PhRvD..62j3008F,2005A&A...433L..53S}.
Therefore, helicity, as well as amplitude, can be an important probe
of the origin of primordial magnetic fields.  

Since magnetic helicity is a conserved quantity in the limit of
infinite conductivity in the standard MHD theory, it should remain
zero if there was no helicity initially in the early universe.  To
generate cosmological magnetic fields with non-zero helicity,
parity violating processes would be necessary, often related with
CP violation of fundamental particle interactions.  This will lead us
to the conclusion that helicity can not be generated if magnetic
fields were generated through
density perturbations where the standard Compton scatterings are the
only relevant particle interactions.  In fact, we can explicitly show
that magnetic fields generated from cosmological perturbation
discussed in this paper have no magnetic helicity as follows.

Formally, the magnetic helicity density is defined as
\begin{equation}
{\cal H} = \frac{1}{V}\int d^3x  \vec{A}\cdot \vec{B}~,
\end{equation}
where $\vec{A}$ is the vector potential and $V$ is the normalization volume. Similar to the way to define
the magnetic field 
spectrum $S(k)$, the helicity spectrum $P_H(k)$ can be defined as
\begin{equation}
 \left<A_i(\vec{k}) B^\ast_j(\vec{k'})\right> =
  \delta^{(3)}(\vec{k}-\vec{k'}) P_{ij}(\hat{k}) \frac{P_H(k)}{2k}~,
\label{eq:P_H}
\end{equation}
where $P_{ij}(\hat{k})=\delta_{ij}-\hat{k}_i \hat{k}_j$ is the
projection tensor$^\dag$.  \footnotetext[2]{In general, $P_{ij}$ is
not necessarily a symmetric tensor. In fact, three conditions we put
on magnetic fields, i.e., reality, divergence-less, and statistical
homogeneity and isotropy require that $P_{ij}$ should have the form:
$P_{ij}(\hat
k)=f(k)\left[\delta_{ij}-\hat{k}^i\hat{k}^j\right]+ig(k)\epsilon_{ijl}\hat{k}^l$
\cite{1977AmJPh..45.1010M,2002PhRvD..65h3502P}.  However, as long as
magnetic helicity is concerned, antisymmetric part in
Eq. (\ref{eq:P_H}) is irrelevant because helicity is defined as an
inner product.}  Note that $P_{ij}$ satisfies the divergence-less
condition of magnetic fields, $k^i P_{ij}=0$. Remembering that $B_i=i
\epsilon_{ijk} k_j A_k$, Eq. (\ref{eq:P_H}) can be put together with
magnetic field spectrum $S(k)$ as
\begin{equation}
\left< B_i(\vec{k}) B^\ast_j(\vec{k'})\right>=
 \delta^{(3)}(\vec{k}-\vec{k'})
\left[
P_{ij}(\hat{k})\frac{S(k)}{2}+i\epsilon_{ijl}\hat{k}^l\frac{P_H(k)}{2}
\right]~.
\label{eq:BiBj}
\end{equation}
Thus, the antisymmetric part of magnetic field correlations denotes
helicity of magnetic fields.

To see explicitly that magnetic fields generated from density
perturbations have no helicity, let us consider an alternative expression for
Eq. (\ref{eq:BiBj})
\begin{equation}
\left< B_j(\vec{k}) B^\ast_i(\vec{k'})\right>=
 \delta^{(3)}(\vec{k}-\vec{k'})
\left[
P_{ij}(\hat{k})\frac{S(k)}{2}-i\epsilon_{ijl}\hat{k}^l\frac{P_H(k)}{2}
\right]~,
\label{eq:BiBj2}
\end{equation}
and subtract this from Eq.(\ref{eq:BiBj}), which should be zero if there
is no magnetic helicity. The magnetic fields generated from
cosmological perturbations can be symbolically expressed as
\begin{equation}
B_i(\vec{k}) = \epsilon^{iab}\int d^3 \vec{K}\hat{K}_a k_b f(\vec{K},\vec{k})
			      \delta(\vec{k}-\vec{K})\gamma(\vec{K})~,
\end{equation}
where $f(\vec{K},\vec{k})$ is a real function of $\vec{K}$ and
$\vec{k}$ (eq.(\ref{eq:a2Bdot})), and $\delta$ and $\gamma$ are the
time integration of perturbation variables.  Let us now explicitly
evaluate
$\left<B_i(\vec{k})B_j^\ast(\vec{k}^\prime)\right>-\left<B_j(\vec{k})B_i^\ast(\vec{k}^\prime)\right>$
as
\begin{eqnarray}
\left<B_i(\vec{k})B_j^\ast(\vec{k}^\prime)\right>-\left<B_j(\vec{k})B_i^\ast(\vec{k}^\prime)\right>
&=&\delta^{(3)}(\vec{k}-\vec{k}^\prime)\int d^3\vec{K}
P(K)P(|\vec{k}-\vec{K}|)f(\vec{K},\vec{k})\epsilon^{iab}\epsilon^{jlm}
\nonumber \\
&\times&\biggl\{(\Khat_a k_b \Khat_l k_m-\Khat_l k_m \Khat_a
 k_b)f(\vec{K},\vec{k})\gamma^2(K)\delta^2(|\vec{k}-\vec{K}|) \nonumber
 \\
&+&\frac{\left(\Khat_a k_b (k_l-K_l)k_m-\Khat_l k_m
	  (k_a-K_a)k_b\right)}{|\vec{k}-\vec{K}|}f(\vec{k}-\vec{K},\vec{k})\delta(|\vec{k}-\vec{K}|)\gamma(|\vec{k}-\vec{K}|)\gamma(K)\delta(K)\biggr\}\nonumber \\
&=&0~.
\end{eqnarray}
Thus, the helical part of magnetic fields from density perturbations should
vanish. This fact is in marked contrast to the other origins of primordial
magnetic fields in the very early universe (above GeV), where the
magnetic fields must be somewhat helical  due
to the interaction of the magnetic field with a cosmic axion field \cite{2005PhRvD..72l3001C}.

Several ways to detect helicity of magnetic fields have been proposed.
It was shown that primordial helicity can be a new source to induce
parity-odd correlations such as between temperature anisotropy and B-mode
polarization, and that between E-mode and B-mode polarizations in
CMB anisotropies \cite{2002PhRvD..65h3502P,2004PhRvD..69f3006C,2005PhRvD..71j3006K}, which are zero for
fields without any helicity.
Recently, Kahniashvili and Vachaspati proposed that the correlation of
the arrival momenta of the cosmic rays can be used to detect 
helicity of an intervening magnetic field \cite{2006PhRvD..73f3507K}.
Although it is very difficult to detect helicity in the cosmological
magnetic fields, yet it deserves further investigations since detecting
helicity  will shed light on the mystery of the origin of large-scale
magnetic fields. 

\section{discussion and summary}
In this paper we discussed a generation mechanism of magnetic fields
from cosmological density perturbations. Following our previous
papers, we present a detailed formulation for the magnetic field
generation. The key is that photons scatter off only electrons (and
not protons) by Compton scatterings. Electric fields are induced to
prevent charge separation between electrons and protons, and magnetic
fields are generated from these induced electric fields through
Maxwell equations.  We showed that the baryon-photon slip and the
anisotropic stress of photons can generate the magnetic fields if
second order couplings in density perturbations are taken into account
when evaluating Compton scattering terms.  Since these two sources for
magnetic fields naturally arise from density perturbations for a wide
range of wavenumbers, the fields are also naturally generated for a
wide range of scales.  We then gave an analytic interpretation of the
resultant spectrum of magnetic fields within the framework of the
cosmological perturbation theory.  Using the tight-coupling
approximation of baryon-photon plasma, we showed that the fields have
the spectrum $\sqrt{k^3 S(k)}\propto k$ at small scales where the
photon anisotropic stress is the dominant contributor, and thus the
fields become stronger at smaller scales.  A typical amplitude of
magnetic fields is of the order $10^{-24}$ Gauss at 1 Mpc scale.
Magnetic helicity should not be associated with these magnetic
fields. We also found that magnetic fields have a cut-off around
$2\times 10^{-3}$
pc due to the relativistic effects of electrons.

In our previous papers \cite{2005PhRvL..95l1301T,2006Sci...311..827I}
we suggested that the magnetic fields as strong as $10^{-18}$ Gauss
can arise at Mpc scale at recombination, which is larger than the
amplitude presented in this paper.  The main difference from the
previous results comes from the missing scale factor in equation
(\ref{eq:B_dot}). This mistake leads an overall overestimate for the
magnetic field amplitude and a steeper magnetic field spectrum on
small scales, where the fields are mainly created at earlier epochs.
The amplitude obtained here is widely consistent with recent
literatures around a comoving scale of $0.1$ Mpc$^{-1}$
\cite{2005PhRvD..71d3502M,2005MNRAS.363..521G,2006astro.ph..4526S}.
However, there still exists the difference of the amplitude of
magnetic fields at smaller scales. For example, our results show
stronger magnetic fields by the factor of $10^{2}$ than those reported
in \cite{2005PhRvD..71d3502M,2006astro.ph..4526S} at Mpc comoving
scale, and the difference becomes even larger at smaller scales. This
may be because they have evaluated the strength of magnetic fields at
the instance of recombination at which the Silk damping effect have
already erased perturbations on scales smaller than $k\sim 0.1$
Mpc$^{-1}$, while we have solved the evolution equation of magnetic
fields from deep in the radiation epoch, through matter-radiation
equality, to recombination. In ref. \cite{2004APh....21...59B}, it
is also discussed that larger magnetic fields arise at small scales if
one consider the earlier period. It is natural to consider that
magnetic fields generated prior to recombination can survive, 
because the diffusion scale in the highly
conductive primordial plasma due to the Ohmic dissipation is much
smaller than cosmological scales considered here. Therefore, we think
that magnetic fields generated from (nearly scale invariant) density
perturbations should dominate on small scales. We will give further
details about the discrepancy between our work and others elsewhere
soon~\cite{futurework1}.

As discussed in this paper, magnetic fields have a small scale power
up to $2\times 10^{-3}$ pc comoving scale. This scale corresponds to the cosmic
horizon when the temperature of the universe was around MeV. Until
this epoch, thermally created electrons and positrons significantly
suppressed the sources of magnetic fields.  If we assume the scale
invariant spectrum for primordial density perturbations, our results
indicate that magnetic fields as strong as $\sim 10^{-15}$ Gauss can
arise at this scale. Although these
are small compared with the micro Gauss fields observed in present
galaxies, it would be enough for the hydro-dynamical dynamo action to
amplify into the present amplitude during the structure formation
until today \cite{1999PhRvD..60b1301D}.
Of course, if density perturbations have a blue spectrum, or
they are anomalously large at small scales \cite{2006PhRvD..74d3525K}
larger magnetic fields will be generated.

Although cosmologically generated magnetic fields appear sufficient for
'seed fields' of galactic magnetic fields, it is still unclear whether
they can act as seed magnetic fields for galaxy clusters. In most 
astrophysical objects, such as disk galaxies and stars, differential
rotation is important for amplifying and sustaining their own magnetic
fields. However, galaxy clusters have little rotation, and therefore,
another mechanisms will be necessary.
It is recently argued that cluster plasmas threaded by weak magnetic
fields are subject to very fast growing instabilities, and this instability
happens once the ions are magnetized \cite{2005ApJ...629..139S}. The
magnetized condition  
corresponds to the magnetic field amplitude of $10^{-18}$ G for typical
cluster parameters, which is a little larger than that of
cosmologically generated seed fields which we found here. 
Furthermore, the fields amplified
by plasma instabilities have a rather small reversal scale (typically
$10^4$ km to 10pc \cite{2006AN....327..599S}), while
the observational data suggest that the 
typical reversal scale is $\sim 1$ kpc. Therefore we can not conclude
that magnetic fields in clusters of galaxies can directly arise from the
seed fields generated from density perturbations in the early universe.

To observe these seed magnetic fields in a direct manner is
interesting but very challenging. The magnetic fields at recombination
should leave an imprint on CMB photons through Faraday
rotation. Faraday rotation can then create B-mode polarization from
the dominant E-mode polarization which associates with (scalar)
density perturbations. Furthermore, sufficiently large magnetic fields
can source the vector mode perturbations on baryon fluid by
themselves, which then induce B-mode polarization pattern in CMB
\cite{2005ApJ...625L...1Y,2004PhRvD..70d3011L}.  By analyzing
polarization patterns at large scales one can in principle get rid of
information about the magnetic fields. Observation forecast predicts
that the fields as week as $10^{-10}$ G will be detected by future
missions such as Planck, which is however still out of
reach of the weak magnetic 'seed' field created from density
perturbations discussed here.

Another promising way to detect the weakest 'seed' magnetic fields
would be to measure the time-delayed emissions from the gamma-ray
bursts. The idea was originally proposed by Plaga
\cite{1995Natur.374..430P}, and there the author concluded that
magnetic fields as week as $10^{-24}$ Gauss in void regions can be
probed by detecting time-delayed events of GeV-TeV photons. Of course
one may claim that it is highly uncertain whether magnetic fields in
void regions purely consist of the remnant seed fields created in the
early universe. Interestingly, it is argued that large volume of void
regions can escape from any astrophysical activities such as galactic
winds \cite{2006MNRAS.tmp..602B}.  Therefore, it is likely that
one can probe primordial magnetic fields by measuring void fields
through gamma-ray photons.  Unfortunately, it would be impossible to
reach the field as weak as $10^{-24}$ Gauss in practice, because there
exist another effects to cause time delays to high energy gamma-ray
photons. Subsequent studies have shown that magnetic fields strength
as strong as $10^{-19}$ G would be necessary for time-delays by
magnetic deflection to be longer than those caused by angular
spreading \cite{2004ApJ...613.1072R}. Therefore, it appears that an
amplification of 'seed' magnetic fields at recombination by at least
$\sim$ 100 will be necessary for the fields to lead any observational
signatures in gamma-ray photons. Clearly it will be of great interest
to study the evolution of magnetic fields in the void regions from
recombination to the present universe.

\acknowledgements 
KI and KT are supported by a Grant-in-Aid for the
Japan Society for the Promotion of Science Fellows and are research
fellows of the Japan Society for the Promotion of Science. N.S. is
supported by a Grant-in-Aid for Scientific Research from the Japanese
Ministry of Education (No. 17540276).  We would like to thank
T. K. Suzuki, M. Hattori and M. Takahashi for helpful suggestions and
useful discussions.

\appendix

\bibliography{part3}

\begin{thebibliography}{69}
\expandafter\ifx\csname natexlab\endcsname\relax\def\natexlab#1{#1}\fi
\expandafter\ifx\csname bibnamefont\endcsname\relax
  \def\bibnamefont#1{#1}\fi
\expandafter\ifx\csname bibfnamefont\endcsname\relax
  \def\bibfnamefont#1{#1}\fi
\expandafter\ifx\csname citenamefont\endcsname\relax
  \def\citenamefont#1{#1}\fi
\expandafter\ifx\csname url\endcsname\relax
  \def\url#1{\texttt{#1}}\fi
\expandafter\ifx\csname urlprefix\endcsname\relax\def\urlprefix{URL }\fi
\providecommand{\bibinfo}[2]{#2}
\providecommand{\eprint}[2][]{\url{#2}}

\bibitem[{\citenamefont{{Widrow}}(2002)}]{2002RvMP...74..775W}
\bibinfo{author}{\bibfnamefont{L.~M.} \bibnamefont{{Widrow}}},
  \bibinfo{journal}{Reviews of Modern Physics} \textbf{\bibinfo{volume}{74}},
  \bibinfo{pages}{775} (\bibinfo{year}{2002}).

\bibitem[{\citenamefont{{Gazzola} et~al.}(2006)\citenamefont{{Gazzola}, {King},
  {Pearce}, and {Coles}}}]{2006astro.ph.11707G}
\bibinfo{author}{\bibfnamefont{L.}~\bibnamefont{{Gazzola}}},
  \bibinfo{author}{\bibfnamefont{E.~J.} \bibnamefont{{King}}},
  \bibinfo{author}{\bibfnamefont{F.~R.} \bibnamefont{{Pearce}}},
  \bibnamefont{and} \bibinfo{author}{\bibfnamefont{P.}~\bibnamefont{{Coles}}},
  \bibinfo{journal}{ArXiv Astrophysics e-prints}  (\bibinfo{year}{2006}),
  \eprint{astro-ph/0611707, to appear in MNRAS}.

\bibitem[{\citenamefont{{Dolag} et~al.}(2001)\citenamefont{{Dolag},
  {Schindler}, {Govoni}, and {Feretti}}}]{2001A&A...378..777D}
\bibinfo{author}{\bibfnamefont{K.}~\bibnamefont{{Dolag}}},
  \bibinfo{author}{\bibfnamefont{S.}~\bibnamefont{{Schindler}}},
  \bibinfo{author}{\bibfnamefont{F.}~\bibnamefont{{Govoni}}}, \bibnamefont{and}
  \bibinfo{author}{\bibfnamefont{L.}~\bibnamefont{{Feretti}}},
  \bibinfo{journal}{A\&A} \textbf{\bibinfo{volume}{378}}, \bibinfo{pages}{777}
  (\bibinfo{year}{2001}), \eprint{astro-ph/0108485}.

\bibitem[{\citenamefont{Dolag et~al.}(1999)\citenamefont{Dolag, Bartelmann, and
  Lesch}}]{Dolag:2002bw}
\bibinfo{author}{\bibfnamefont{K.}~\bibnamefont{Dolag}},
  \bibinfo{author}{\bibfnamefont{M.}~\bibnamefont{Bartelmann}},
  \bibnamefont{and} \bibinfo{author}{\bibfnamefont{H.}~\bibnamefont{Lesch}},
  \bibinfo{journal}{Astron. Astrophys.} \textbf{\bibinfo{volume}{348}},
  \bibinfo{pages}{351} (\bibinfo{year}{1999}), \eprint{astro-ph/0202272}.

\bibitem[{\citenamefont{{Silk} and {Langer}}(2006)}]{2006MNRAS.371..444S}
\bibinfo{author}{\bibfnamefont{J.}~\bibnamefont{{Silk}}} \bibnamefont{and}
  \bibinfo{author}{\bibfnamefont{M.}~\bibnamefont{{Langer}}},
  \bibinfo{journal}{\mnras} \textbf{\bibinfo{volume}{371}},
  \bibinfo{pages}{444} (\bibinfo{year}{2006}), \eprint{astro-ph/0606276}.

\bibitem[{\citenamefont{{Machida} et~al.}(2006)\citenamefont{{Machida},
  {Omukai}, {Matsumoto}, and {Inutsuka}}}]{2006ApJ...647L...1M}
\bibinfo{author}{\bibfnamefont{M.~N.} \bibnamefont{{Machida}}},
  \bibinfo{author}{\bibfnamefont{K.}~\bibnamefont{{Omukai}}},
  \bibinfo{author}{\bibfnamefont{T.}~\bibnamefont{{Matsumoto}}},
  \bibnamefont{and} \bibinfo{author}{\bibfnamefont{S.-i.}
  \bibnamefont{{Inutsuka}}}, \bibinfo{journal}{\apjl}
  \textbf{\bibinfo{volume}{647}}, \bibinfo{pages}{L1} (\bibinfo{year}{2006}),
  \eprint{astro-ph/0605146}.

\bibitem[{\citenamefont{{Davis} et~al.}(1999)\citenamefont{{Davis}, {Lilley},
  and {T{\"o}rnkvist}}}]{1999PhRvD..60b1301D}
\bibinfo{author}{\bibfnamefont{A.-C.} \bibnamefont{{Davis}}},
  \bibinfo{author}{\bibfnamefont{M.}~\bibnamefont{{Lilley}}}, \bibnamefont{and}
  \bibinfo{author}{\bibfnamefont{O.}~\bibnamefont{{T{\"o}rnkvist}}},
  \bibinfo{journal}{\prd} \textbf{\bibinfo{volume}{60}},
  \bibinfo{pages}{021301} (\bibinfo{year}{1999}).

\bibitem[{\citenamefont{{Wolfe} et~al.}(1992)\citenamefont{{Wolfe}, {Lanzetta},
  and {Oren}}}]{1992ApJ...388...17W}
\bibinfo{author}{\bibfnamefont{A.~M.} \bibnamefont{{Wolfe}}},
  \bibinfo{author}{\bibfnamefont{K.~M.} \bibnamefont{{Lanzetta}}},
  \bibnamefont{and} \bibinfo{author}{\bibfnamefont{A.~L.}
  \bibnamefont{{Oren}}}, \bibinfo{journal}{\apj}
  \textbf{\bibinfo{volume}{388}}, \bibinfo{pages}{17} (\bibinfo{year}{1992}).

\bibitem[{\citenamefont{{Kronberg} et~al.}(1992)\citenamefont{{Kronberg},
  {Perry}, and {Zukowski}}}]{1992ApJ...387..528K}
\bibinfo{author}{\bibfnamefont{P.~P.} \bibnamefont{{Kronberg}}},
  \bibinfo{author}{\bibfnamefont{J.~J.} \bibnamefont{{Perry}}},
  \bibnamefont{and} \bibinfo{author}{\bibfnamefont{E.~L.~H.}
  \bibnamefont{{Zukowski}}}, \bibinfo{journal}{\apj}
  \textbf{\bibinfo{volume}{387}}, \bibinfo{pages}{528} (\bibinfo{year}{1992}).

\bibitem[{\citenamefont{{Biermann} and {Schl{\"u}ter}}(1951)}]{Biermann50}
\bibinfo{author}{\bibfnamefont{L.}~\bibnamefont{{Biermann}}} \bibnamefont{and}
  \bibinfo{author}{\bibfnamefont{A.}~\bibnamefont{{Schl{\"u}ter}}},
  \bibinfo{journal}{Physical Review} \textbf{\bibinfo{volume}{82}},
  \bibinfo{pages}{863} (\bibinfo{year}{1951}).

\bibitem[{\citenamefont{{Kemp}}(1982)}]{1982PASP...94..627K}
\bibinfo{author}{\bibfnamefont{J.~C.} \bibnamefont{{Kemp}}},
  \bibinfo{journal}{Publ. Astron. Soc. Pac.} \textbf{\bibinfo{volume}{94}},
  \bibinfo{pages}{627} (\bibinfo{year}{1982}).

\bibitem[{\citenamefont{{Miranda} et~al.}(1998)\citenamefont{{Miranda},
  {Opher}, and {Opher}}}]{1998MNRAS.301..547M}
\bibinfo{author}{\bibfnamefont{O.~D.} \bibnamefont{{Miranda}}},
  \bibinfo{author}{\bibfnamefont{M.}~\bibnamefont{{Opher}}}, \bibnamefont{and}
  \bibinfo{author}{\bibfnamefont{R.}~\bibnamefont{{Opher}}},
  \bibinfo{journal}{\mnras} \textbf{\bibinfo{volume}{301}},
  \bibinfo{pages}{547} (\bibinfo{year}{1998}).

\bibitem[{\citenamefont{Hanayama et~al.}(2005)}]{Hanayama05}
\bibinfo{author}{\bibfnamefont{H.}~\bibnamefont{Hanayama}}
  \bibnamefont{et~al.}, \bibinfo{journal}{Astrophys. J.}
  \textbf{\bibinfo{volume}{633}}, \bibinfo{pages}{941} (\bibinfo{year}{2005}),
  \eprint{astro-ph/0501538}.

\bibitem[{\citenamefont{{Davies} and {Widrow}}(2000)}]{2000ApJ...540..755D}
\bibinfo{author}{\bibfnamefont{G.}~\bibnamefont{{Davies}}} \bibnamefont{and}
  \bibinfo{author}{\bibfnamefont{L.~M.} \bibnamefont{{Widrow}}},
  \bibinfo{journal}{\apj} \textbf{\bibinfo{volume}{540}}, \bibinfo{pages}{755}
  (\bibinfo{year}{2000}).

\bibitem[{\citenamefont{{Kulsrud} et~al.}(1997)\citenamefont{{Kulsrud}, {Cen},
  {Ostriker}, and {Ryu}}}]{1997ApJ...480..481K}
\bibinfo{author}{\bibfnamefont{R.~M.} \bibnamefont{{Kulsrud}}},
  \bibinfo{author}{\bibfnamefont{R.}~\bibnamefont{{Cen}}},
  \bibinfo{author}{\bibfnamefont{J.~P.} \bibnamefont{{Ostriker}}},
  \bibnamefont{and} \bibinfo{author}{\bibfnamefont{D.}~\bibnamefont{{Ryu}}},
  \bibinfo{journal}{\apj} \textbf{\bibinfo{volume}{480}}, \bibinfo{pages}{481}
  (\bibinfo{year}{1997}).

\bibitem[{\citenamefont{{Gnedin} et~al.}(2000)\citenamefont{{Gnedin},
  {Ferrara}, and {Zweibel}}}]{2000ApJ...539..505G}
\bibinfo{author}{\bibfnamefont{N.~Y.} \bibnamefont{{Gnedin}}},
  \bibinfo{author}{\bibfnamefont{A.}~\bibnamefont{{Ferrara}}},
  \bibnamefont{and} \bibinfo{author}{\bibfnamefont{E.~G.}
  \bibnamefont{{Zweibel}}}, \bibinfo{journal}{\apj}
  \textbf{\bibinfo{volume}{539}}, \bibinfo{pages}{505} (\bibinfo{year}{2000}).

\bibitem[{\citenamefont{{Fujita} and {Kato}}(2005)}]{2005MNRAS.364..247F}
\bibinfo{author}{\bibfnamefont{Y.}~\bibnamefont{{Fujita}}} \bibnamefont{and}
  \bibinfo{author}{\bibfnamefont{T.~N.} \bibnamefont{{Kato}}},
  \bibinfo{journal}{\mnras} \textbf{\bibinfo{volume}{364}},
  \bibinfo{pages}{247} (\bibinfo{year}{2005}).

\bibitem[{\citenamefont{Medvedev et~al.}(2005)\citenamefont{Medvedev, Silva,
  and Kamionkowski}}]{Medvedev:2005ep}
\bibinfo{author}{\bibfnamefont{M.~V.} \bibnamefont{Medvedev}},
  \bibinfo{author}{\bibfnamefont{L.~O.} \bibnamefont{Silva}}, \bibnamefont{and}
  \bibinfo{author}{\bibfnamefont{M.}~\bibnamefont{Kamionkowski}}
  (\bibinfo{year}{2005}), \eprint{astro-ph/0512079}.

\bibitem[{\citenamefont{Ratra}(1992)}]{Ratra:1991bn}
\bibinfo{author}{\bibfnamefont{B.}~\bibnamefont{Ratra}},
  \bibinfo{journal}{Astrophys. J.} \textbf{\bibinfo{volume}{391}},
  \bibinfo{pages}{L1} (\bibinfo{year}{1992}).

\bibitem[{\citenamefont{Bamba and Yokoyama}(2004)}]{Bamba:2003av}
\bibinfo{author}{\bibfnamefont{K.}~\bibnamefont{Bamba}} \bibnamefont{and}
  \bibinfo{author}{\bibfnamefont{J.}~\bibnamefont{Yokoyama}},
  \bibinfo{journal}{Phys. Rev.} \textbf{\bibinfo{volume}{D69}},
  \bibinfo{pages}{043507} (\bibinfo{year}{2004}), \eprint{astro-ph/0310824}.

\bibitem[{\citenamefont{{Prokopec} and {Puchwein}}(2004)}]{2004PhRvD..70d3004P}
\bibinfo{author}{\bibfnamefont{T.}~\bibnamefont{{Prokopec}}} \bibnamefont{and}
  \bibinfo{author}{\bibfnamefont{E.}~\bibnamefont{{Puchwein}}},
  \bibinfo{journal}{\prd} \textbf{\bibinfo{volume}{70}},
  \bibinfo{pages}{043004} (\bibinfo{year}{2004}).

\bibitem[{\citenamefont{Turner and Widrow}(1988)}]{Turner:1987bw}
\bibinfo{author}{\bibfnamefont{M.~S.} \bibnamefont{Turner}} \bibnamefont{and}
  \bibinfo{author}{\bibfnamefont{L.~M.} \bibnamefont{Widrow}},
  \bibinfo{journal}{Phys. Rev.} \textbf{\bibinfo{volume}{D37}},
  \bibinfo{pages}{2743} (\bibinfo{year}{1988}).

\bibitem[{\citenamefont{Caprini and Durrer}(2002)}]{Caprini:2001nb}
\bibinfo{author}{\bibfnamefont{C.}~\bibnamefont{Caprini}} \bibnamefont{and}
  \bibinfo{author}{\bibfnamefont{R.}~\bibnamefont{Durrer}},
  \bibinfo{journal}{Phys. Rev.} \textbf{\bibinfo{volume}{D65}},
  \bibinfo{pages}{023517} (\bibinfo{year}{2002}), \eprint{astro-ph/0106244}.

\bibitem[{\citenamefont{{Caprini} and {Durrer}}(2005)}]{2005PhRvD..72h8301C}
\bibinfo{author}{\bibfnamefont{C.}~\bibnamefont{{Caprini}}} \bibnamefont{and}
  \bibinfo{author}{\bibfnamefont{R.}~\bibnamefont{{Durrer}}},
  \bibinfo{journal}{\prd} \textbf{\bibinfo{volume}{72}},
  \bibinfo{pages}{088301} (\bibinfo{year}{2005}).

\bibitem[{\citenamefont{{Harrison}}(1970)}]{1970MNRAS.147..279H}
\bibinfo{author}{\bibfnamefont{E.~R.} \bibnamefont{{Harrison}}},
  \bibinfo{journal}{\mnras} \textbf{\bibinfo{volume}{147}},
  \bibinfo{pages}{279} (\bibinfo{year}{1970}).

\bibitem[{\citenamefont{Hogan}(2000)}]{Hogan:2000gv}
\bibinfo{author}{\bibfnamefont{C.~J.} \bibnamefont{Hogan}}
  (\bibinfo{year}{2000}), \eprint{astro-ph/0005380}.

\bibitem[{\citenamefont{{Matarrese} et~al.}(2005)\citenamefont{{Matarrese},
  {Mollerach}, {Notari}, and {Riotto}}}]{2005PhRvD..71d3502M}
\bibinfo{author}{\bibfnamefont{S.}~\bibnamefont{{Matarrese}}},
  \bibinfo{author}{\bibfnamefont{S.}~\bibnamefont{{Mollerach}}},
  \bibinfo{author}{\bibfnamefont{A.}~\bibnamefont{{Notari}}}, \bibnamefont{and}
  \bibinfo{author}{\bibfnamefont{A.}~\bibnamefont{{Riotto}}},
  \bibinfo{journal}{\prd} \textbf{\bibinfo{volume}{71}},
  \bibinfo{pages}{043502} (\bibinfo{year}{2005}).

\bibitem[{\citenamefont{{Berezhiani} and {Dolgov}}(2004)}]{2004APh....21...59B}
\bibinfo{author}{\bibfnamefont{Z.}~\bibnamefont{{Berezhiani}}}
  \bibnamefont{and} \bibinfo{author}{\bibfnamefont{A.~D.}
  \bibnamefont{{Dolgov}}}, \bibinfo{journal}{Astroparticle Physics}
  \textbf{\bibinfo{volume}{21}}, \bibinfo{pages}{59} (\bibinfo{year}{2004}).

\bibitem[{\citenamefont{{Gopal} and {Sethi}}(2005)}]{2005MNRAS.363..521G}
\bibinfo{author}{\bibfnamefont{R.}~\bibnamefont{{Gopal}}} \bibnamefont{and}
  \bibinfo{author}{\bibfnamefont{S.~K.} \bibnamefont{{Sethi}}},
  \bibinfo{journal}{\mnras} \textbf{\bibinfo{volume}{363}},
  \bibinfo{pages}{521} (\bibinfo{year}{2005}).

\bibitem[{\citenamefont{{Takahashi} et~al.}(2005)\citenamefont{{Takahashi},
  {Ichiki}, {Ohno}, and {Hanayama}}}]{2005PhRvL..95l1301T}
\bibinfo{author}{\bibfnamefont{K.}~\bibnamefont{{Takahashi}}},
  \bibinfo{author}{\bibfnamefont{K.}~\bibnamefont{{Ichiki}}},
  \bibinfo{author}{\bibfnamefont{H.}~\bibnamefont{{Ohno}}}, \bibnamefont{and}
  \bibinfo{author}{\bibfnamefont{H.}~\bibnamefont{{Hanayama}}},
  \bibinfo{journal}{Physical Review Letters} \textbf{\bibinfo{volume}{95}},
  \bibinfo{pages}{121301} (\bibinfo{year}{2005}).

\bibitem[{\citenamefont{{Ichiki} et~al.}(2006)\citenamefont{{Ichiki},
  {Takahashi}, {Ohno}, {Hanayama}, and {Sugiyama}}}]{2006Sci...311..827I}
\bibinfo{author}{\bibfnamefont{K.}~\bibnamefont{{Ichiki}}},
  \bibinfo{author}{\bibfnamefont{K.}~\bibnamefont{{Takahashi}}},
  \bibinfo{author}{\bibfnamefont{H.}~\bibnamefont{{Ohno}}},
  \bibinfo{author}{\bibfnamefont{H.}~\bibnamefont{{Hanayama}}},
  \bibnamefont{and}
  \bibinfo{author}{\bibfnamefont{N.}~\bibnamefont{{Sugiyama}}},
  \bibinfo{journal}{Science} \textbf{\bibinfo{volume}{311}},
  \bibinfo{pages}{827} (\bibinfo{year}{2006}).

\bibitem[{\citenamefont{{Hu}}(1995)}]{1995PhDT..........H}
\bibinfo{author}{\bibfnamefont{W.~T.} \bibnamefont{{Hu}}},
  \bibinfo{journal}{Ph.D.~Thesis}  (\bibinfo{year}{1995}).

\bibitem[{\citenamefont{{Dodelson} and {Jubas}}(1995)}]{1995ApJ...439..503D}
\bibinfo{author}{\bibfnamefont{S.}~\bibnamefont{{Dodelson}}} \bibnamefont{and}
  \bibinfo{author}{\bibfnamefont{J.~M.} \bibnamefont{{Jubas}}},
  \bibinfo{journal}{\apj} \textbf{\bibinfo{volume}{439}}, \bibinfo{pages}{503}
  (\bibinfo{year}{1995}).

\bibitem[{\citenamefont{{Bartolo} et~al.}(2006)\citenamefont{{Bartolo},
  {Matarrese}, and {Riotto}}}]{2006astro.ph..4416B}
\bibinfo{author}{\bibfnamefont{N.}~\bibnamefont{{Bartolo}}},
  \bibinfo{author}{\bibfnamefont{S.}~\bibnamefont{{Matarrese}}},
  \bibnamefont{and} \bibinfo{author}{\bibfnamefont{A.}~\bibnamefont{{Riotto}}},
  \bibinfo{journal}{ArXiv Astrophysics e-prints}  (\bibinfo{year}{2006}),
  \eprint{astro-ph/0604416}.

\bibitem[{\citenamefont{{Subramanian} et~al.}(1994)\citenamefont{{Subramanian},
  {Narasimha}, and {Chitre}}}]{1994MNRAS.271L..15S}
\bibinfo{author}{\bibfnamefont{K.}~\bibnamefont{{Subramanian}}},
  \bibinfo{author}{\bibfnamefont{D.}~\bibnamefont{{Narasimha}}},
  \bibnamefont{and} \bibinfo{author}{\bibfnamefont{S.~M.}
  \bibnamefont{{Chitre}}}, \bibinfo{journal}{\mnras}
  \textbf{\bibinfo{volume}{271}}, \bibinfo{pages}{L15+} (\bibinfo{year}{1994}).

\bibitem[{\citenamefont{{Jedamzik} et~al.}(1998)\citenamefont{{Jedamzik},
  {Katalini{\'c}}, and {Olinto}}}]{1998PhRvD..57.3264J}
\bibinfo{author}{\bibfnamefont{K.}~\bibnamefont{{Jedamzik}}},
  \bibinfo{author}{\bibfnamefont{V.}~\bibnamefont{{Katalini{\'c}}}},
  \bibnamefont{and} \bibinfo{author}{\bibfnamefont{A.~V.}
  \bibnamefont{{Olinto}}}, \bibinfo{journal}{\prd}
  \textbf{\bibinfo{volume}{57}}, \bibinfo{pages}{3264} (\bibinfo{year}{1998}).

\bibitem[{\citenamefont{{Seljak} et~al.}(2005)\citenamefont{{Seljak},
  {Makarov}, {McDonald}, {Anderson}, {Bahcall}, {Brinkmann}, {Burles}, {Cen},
  {Doi}, {Gunn} et~al.}}]{2005PhRvD..71j3515S}
\bibinfo{author}{\bibfnamefont{U.}~\bibnamefont{{Seljak}}},
  \bibinfo{author}{\bibfnamefont{A.}~\bibnamefont{{Makarov}}},
  \bibinfo{author}{\bibfnamefont{P.}~\bibnamefont{{McDonald}}},
  \bibinfo{author}{\bibfnamefont{S.~F.} \bibnamefont{{Anderson}}},
  \bibinfo{author}{\bibfnamefont{N.~A.} \bibnamefont{{Bahcall}}},
  \bibinfo{author}{\bibfnamefont{J.}~\bibnamefont{{Brinkmann}}},
  \bibinfo{author}{\bibfnamefont{S.}~\bibnamefont{{Burles}}},
  \bibinfo{author}{\bibfnamefont{R.}~\bibnamefont{{Cen}}},
  \bibinfo{author}{\bibfnamefont{M.}~\bibnamefont{{Doi}}},
  \bibinfo{author}{\bibfnamefont{J.~E.} \bibnamefont{{Gunn}}},
  \bibnamefont{et~al.}, \bibinfo{journal}{\prd} \textbf{\bibinfo{volume}{71}},
  \bibinfo{pages}{103515} (\bibinfo{year}{2005}).

\bibitem[{\citenamefont{{Ma} and {Bertschinger}}(1995)}]{1995ApJ...455....7M}
\bibinfo{author}{\bibfnamefont{C.-P.} \bibnamefont{{Ma}}} \bibnamefont{and}
  \bibinfo{author}{\bibfnamefont{E.}~\bibnamefont{{Bertschinger}}},
  \bibinfo{journal}{\apj} \textbf{\bibinfo{volume}{455}}, \bibinfo{pages}{7}
  (\bibinfo{year}{1995}).

\bibitem[{\citenamefont{Lifshitz}(1946)}]{Lifshitz:1945du}
\bibinfo{author}{\bibfnamefont{E.}~\bibnamefont{Lifshitz}},
  \bibinfo{journal}{J. Phys. (USSR)} \textbf{\bibinfo{volume}{10}},
  \bibinfo{pages}{116} (\bibinfo{year}{1946}).

\bibitem[{\citenamefont{{Landau} and {Lifshitz}}(1971)}]{1971ctf..book.....L}
\bibinfo{author}{\bibfnamefont{L.~D.} \bibnamefont{{Landau}}} \bibnamefont{and}
  \bibinfo{author}{\bibfnamefont{E.~M.} \bibnamefont{{Lifshitz}}},
  \emph{\bibinfo{title}{{The classical theory of fields}}}
  (\bibinfo{publisher}{Course of theoretical physics - Pergamon International
  Library of Science, Technology, Engineering and Social Studies, Oxford:
  Pergamon Press, 1971, 3rd rev.~engl.~edition}, \bibinfo{year}{1971}).

\bibitem[{\citenamefont{{Kodama} and {Sasaki}}(1984)}]{1984PThPS..78....1K}
\bibinfo{author}{\bibfnamefont{H.}~\bibnamefont{{Kodama}}} \bibnamefont{and}
  \bibinfo{author}{\bibfnamefont{M.}~\bibnamefont{{Sasaki}}},
  \bibinfo{journal}{Progress of Theoretical Physics Supplement}
  \textbf{\bibinfo{volume}{78}}, \bibinfo{pages}{1} (\bibinfo{year}{1984}).

\bibitem[{\citenamefont{{Mukhanov} et~al.}(1992)\citenamefont{{Mukhanov},
  {Feldman}, and {Brandenberger}}}]{1992PhR...215..203M}
\bibinfo{author}{\bibfnamefont{V.~F.} \bibnamefont{{Mukhanov}}},
  \bibinfo{author}{\bibfnamefont{H.~A.} \bibnamefont{{Feldman}}},
  \bibnamefont{and} \bibinfo{author}{\bibfnamefont{R.~H.}
  \bibnamefont{{Brandenberger}}}, \bibinfo{journal}{Phys. Rep.}
  \textbf{\bibinfo{volume}{215}}, \bibinfo{pages}{203} (\bibinfo{year}{1992}).

\bibitem[{\citenamefont{{Peebles} and {Yu}}(1970)}]{1970ApJ...162..815P}
\bibinfo{author}{\bibfnamefont{P.~J.~E.} \bibnamefont{{Peebles}}}
  \bibnamefont{and} \bibinfo{author}{\bibfnamefont{J.~T.} \bibnamefont{{Yu}}},
  \bibinfo{journal}{\apj} \textbf{\bibinfo{volume}{162}}, \bibinfo{pages}{815}
  (\bibinfo{year}{1970}).

\bibitem[{\citenamefont{{Bond} and {Szalay}}(1983)}]{1983ApJ...274..443B}
\bibinfo{author}{\bibfnamefont{J.~R.} \bibnamefont{{Bond}}} \bibnamefont{and}
  \bibinfo{author}{\bibfnamefont{A.~S.} \bibnamefont{{Szalay}}},
  \bibinfo{journal}{\apj} \textbf{\bibinfo{volume}{274}}, \bibinfo{pages}{443}
  (\bibinfo{year}{1983}).

\bibitem[{\citenamefont{{Bond} and {Efstathiou}}(1984)}]{1984ApJ...285L..45B}
\bibinfo{author}{\bibfnamefont{J.~R.} \bibnamefont{{Bond}}} \bibnamefont{and}
  \bibinfo{author}{\bibfnamefont{G.}~\bibnamefont{{Efstathiou}}},
  \bibinfo{journal}{\apjl} \textbf{\bibinfo{volume}{285}}, \bibinfo{pages}{L45}
  (\bibinfo{year}{1984}).

\bibitem[{\citenamefont{{Hu} and {Sugiyama}}(1995)}]{1995ApJ...444..489H}
\bibinfo{author}{\bibfnamefont{W.}~\bibnamefont{{Hu}}} \bibnamefont{and}
  \bibinfo{author}{\bibfnamefont{N.}~\bibnamefont{{Sugiyama}}},
  \bibinfo{journal}{\apj} \textbf{\bibinfo{volume}{444}}, \bibinfo{pages}{489}
  (\bibinfo{year}{1995}).

\bibitem[{\citenamefont{{Seljak} and
  {Zaldarriaga}}(1996)}]{1996ApJ...469..437S}
\bibinfo{author}{\bibfnamefont{U.}~\bibnamefont{{Seljak}}} \bibnamefont{and}
  \bibinfo{author}{\bibfnamefont{M.}~\bibnamefont{{Zaldarriaga}}},
  \bibinfo{journal}{\apj} \textbf{\bibinfo{volume}{469}}, \bibinfo{pages}{437}
  (\bibinfo{year}{1996}), \eprint{astro-ph/9603033}.

\bibitem[{\citenamefont{Lewis et~al.}(2000)\citenamefont{Lewis, Challinor, and
  Lasenby}}]{Lewis:1999bs}
\bibinfo{author}{\bibfnamefont{A.}~\bibnamefont{Lewis}},
  \bibinfo{author}{\bibfnamefont{A.}~\bibnamefont{Challinor}},
  \bibnamefont{and} \bibinfo{author}{\bibfnamefont{A.}~\bibnamefont{Lasenby}},
  \bibinfo{journal}{Astrophys. J.} \textbf{\bibinfo{volume}{538}},
  \bibinfo{pages}{473} (\bibinfo{year}{2000}), \eprint{astro-ph/9911177}.

\bibitem[{\citenamefont{{Silk}}(1968)}]{1968ApJ...151..459S}
\bibinfo{author}{\bibfnamefont{J.}~\bibnamefont{{Silk}}},
  \bibinfo{journal}{\apj} \textbf{\bibinfo{volume}{151}}, \bibinfo{pages}{459}
  (\bibinfo{year}{1968}).

\bibitem[{\citenamefont{{Takahashi} et~al.}(2006)\citenamefont{{Takahashi},
  {Ichiki}, and {Sugiyama}}}]{futurework1}
\bibinfo{author}{\bibfnamefont{K.}~\bibnamefont{{Takahashi}}},
  \bibinfo{author}{\bibfnamefont{K.}~\bibnamefont{{Ichiki}}}, \bibnamefont{and}
  \bibinfo{author}{\bibfnamefont{N.}~\bibnamefont{{Sugiyama}}},
  \bibinfo{journal}{in preparation,}  (\bibinfo{year}{2006}).

\bibitem[{\citenamefont{{Biskamp}}(1993)}]{1993noma.book.....B}
\bibinfo{author}{\bibfnamefont{D.}~\bibnamefont{{Biskamp}}},
  \emph{\bibinfo{title}{{Nonlinear magnetohydrodynamics}}}
  (\bibinfo{publisher}{Cambridge Monographs on Plasma Physics, Cambridge
  [England]; New York, NY: Cambridge University Press, |c1993},
  \bibinfo{year}{1993}).

\bibitem[{\citenamefont{{Vachaspati}}(2001)}]{2001PhRvL..87y1302V}
\bibinfo{author}{\bibfnamefont{T.}~\bibnamefont{{Vachaspati}}},
  \bibinfo{journal}{Physical Review Letters} \textbf{\bibinfo{volume}{87}},
  \bibinfo{pages}{251302} (\bibinfo{year}{2001}), \eprint{astro-ph/0101261}.

\bibitem[{\citenamefont{{Field} and {Carroll}}(2000)}]{2000PhRvD..62j3008F}
\bibinfo{author}{\bibfnamefont{G.~B.} \bibnamefont{{Field}}} \bibnamefont{and}
  \bibinfo{author}{\bibfnamefont{S.~M.} \bibnamefont{{Carroll}}},
  \bibinfo{journal}{\prd} \textbf{\bibinfo{volume}{62}},
  \bibinfo{pages}{103008} (\bibinfo{year}{2000}), \eprint{astro-ph/9811206}.

\bibitem[{\citenamefont{{Semikoz} and {Sokoloff}}(2005)}]{2005A&A...433L..53S}
\bibinfo{author}{\bibfnamefont{V.~B.} \bibnamefont{{Semikoz}}}
  \bibnamefont{and}
  \bibinfo{author}{\bibfnamefont{D.}~\bibnamefont{{Sokoloff}}},
  \bibinfo{journal}{A\&A} \textbf{\bibinfo{volume}{433}}, \bibinfo{pages}{L53}
  (\bibinfo{year}{2005}), \eprint{astro-ph/0411496}.

\bibitem[{\citenamefont{{Monin} et~al.}(1977)\citenamefont{{Monin}, {Yaglom},
  and {Ablow}}}]{1977AmJPh..45.1010M}
\bibinfo{author}{\bibfnamefont{A.~S.} \bibnamefont{{Monin}}},
  \bibinfo{author}{\bibfnamefont{A.~M.} \bibnamefont{{Yaglom}}},
  \bibnamefont{and} \bibinfo{author}{\bibfnamefont{C.~M.}
  \bibnamefont{{Ablow}}}, \bibinfo{journal}{American Journal of Physics}
  \textbf{\bibinfo{volume}{45}}, \bibinfo{pages}{1010} (\bibinfo{year}{1977}).

\bibitem[{\citenamefont{{Pogosian} et~al.}(2002)\citenamefont{{Pogosian},
  {Vachaspati}, and {Winitzki}}}]{2002PhRvD..65h3502P}
\bibinfo{author}{\bibfnamefont{L.}~\bibnamefont{{Pogosian}}},
  \bibinfo{author}{\bibfnamefont{T.}~\bibnamefont{{Vachaspati}}},
  \bibnamefont{and}
  \bibinfo{author}{\bibfnamefont{S.}~\bibnamefont{{Winitzki}}},
  \bibinfo{journal}{\prd} \textbf{\bibinfo{volume}{65}},
  \bibinfo{pages}{083502} (\bibinfo{year}{2002}).

\bibitem[{\citenamefont{{Campanelli} and
  {Giannotti}}(2005)}]{2005PhRvD..72l3001C}
\bibinfo{author}{\bibfnamefont{L.}~\bibnamefont{{Campanelli}}}
  \bibnamefont{and}
  \bibinfo{author}{\bibfnamefont{M.}~\bibnamefont{{Giannotti}}},
  \bibinfo{journal}{\prd} \textbf{\bibinfo{volume}{72}},
  \bibinfo{pages}{123001} (\bibinfo{year}{2005}).

\bibitem[{\citenamefont{{Caprini} et~al.}(2004)\citenamefont{{Caprini},
  {Durrer}, and {Kahniashvili}}}]{2004PhRvD..69f3006C}
\bibinfo{author}{\bibfnamefont{C.}~\bibnamefont{{Caprini}}},
  \bibinfo{author}{\bibfnamefont{R.}~\bibnamefont{{Durrer}}}, \bibnamefont{and}
  \bibinfo{author}{\bibfnamefont{T.}~\bibnamefont{{Kahniashvili}}},
  \bibinfo{journal}{\prd} \textbf{\bibinfo{volume}{69}},
  \bibinfo{pages}{063006} (\bibinfo{year}{2004}), \eprint{astro-ph/0304556}.

\bibitem[{\citenamefont{{Kahniashvili} and
  {Ratra}}(2005)}]{2005PhRvD..71j3006K}
\bibinfo{author}{\bibfnamefont{T.}~\bibnamefont{{Kahniashvili}}}
  \bibnamefont{and} \bibinfo{author}{\bibfnamefont{B.}~\bibnamefont{{Ratra}}},
  \bibinfo{journal}{\prd} \textbf{\bibinfo{volume}{71}},
  \bibinfo{pages}{103006} (\bibinfo{year}{2005}), \eprint{astro-ph/0503709}.

\bibitem[{\citenamefont{{Kahniashvili} and
  {Vachaspati}}(2006)}]{2006PhRvD..73f3507K}
\bibinfo{author}{\bibfnamefont{T.}~\bibnamefont{{Kahniashvili}}}
  \bibnamefont{and}
  \bibinfo{author}{\bibfnamefont{T.}~\bibnamefont{{Vachaspati}}},
  \bibinfo{journal}{\prd} \textbf{\bibinfo{volume}{73}},
  \bibinfo{pages}{063507} (\bibinfo{year}{2006}), \eprint{astro-ph/0511373}.

\bibitem[{\citenamefont{{Siegel} and {Fry}}(2006)}]{2006astro.ph..4526S}
\bibinfo{author}{\bibfnamefont{E.~R.} \bibnamefont{{Siegel}}} \bibnamefont{and}
  \bibinfo{author}{\bibfnamefont{J.~N.} \bibnamefont{{Fry}}},
  \bibinfo{journal}{ArXiv Astrophysics e-prints}  (\bibinfo{year}{2006}),
  \eprint{astro-ph/0604526}.

\bibitem[{\citenamefont{{Kawasaki} et~al.}(2006)\citenamefont{{Kawasaki},
  {Takayama}, {Yamaguchi}, and {Yokoyama}}}]{2006PhRvD..74d3525K}
\bibinfo{author}{\bibfnamefont{M.}~\bibnamefont{{Kawasaki}}},
  \bibinfo{author}{\bibfnamefont{T.}~\bibnamefont{{Takayama}}},
  \bibinfo{author}{\bibfnamefont{M.}~\bibnamefont{{Yamaguchi}}},
  \bibnamefont{and}
  \bibinfo{author}{\bibfnamefont{J.}~\bibnamefont{{Yokoyama}}},
  \bibinfo{journal}{\prd} \textbf{\bibinfo{volume}{74}},
  \bibinfo{pages}{043525} (\bibinfo{year}{2006}), \eprint{hep-ph/0605271}.

\bibitem[{\citenamefont{{Schekochihin}
  et~al.}(2005)\citenamefont{{Schekochihin}, {Cowley}, {Kulsrud}, {Hammett},
  and {Sharma}}}]{2005ApJ...629..139S}
\bibinfo{author}{\bibfnamefont{A.~A.} \bibnamefont{{Schekochihin}}},
  \bibinfo{author}{\bibfnamefont{S.~C.} \bibnamefont{{Cowley}}},
  \bibinfo{author}{\bibfnamefont{R.~M.} \bibnamefont{{Kulsrud}}},
  \bibinfo{author}{\bibfnamefont{G.~W.} \bibnamefont{{Hammett}}},
  \bibnamefont{and} \bibinfo{author}{\bibfnamefont{P.}~\bibnamefont{{Sharma}}},
  \bibinfo{journal}{\apj} \textbf{\bibinfo{volume}{629}}, \bibinfo{pages}{139}
  (\bibinfo{year}{2005}), \eprint{astro-ph/0501362}.

\bibitem[{\citenamefont{{Schekochihin} and
  {Cowley}}(2006)}]{2006AN....327..599S}
\bibinfo{author}{\bibfnamefont{A.~A.} \bibnamefont{{Schekochihin}}}
  \bibnamefont{and} \bibinfo{author}{\bibfnamefont{S.~C.}
  \bibnamefont{{Cowley}}}, \bibinfo{journal}{Astronomische Nachrichten}
  \textbf{\bibinfo{volume}{327}}, \bibinfo{pages}{599} (\bibinfo{year}{2006}),
  \eprint{astro-ph/0508535}.

\bibitem[{\citenamefont{{Yamazaki} et~al.}(2005)\citenamefont{{Yamazaki},
  {Ichiki}, and {Kajino}}}]{2005ApJ...625L...1Y}
\bibinfo{author}{\bibfnamefont{D.~G.} \bibnamefont{{Yamazaki}}},
  \bibinfo{author}{\bibfnamefont{K.}~\bibnamefont{{Ichiki}}}, \bibnamefont{and}
  \bibinfo{author}{\bibfnamefont{T.}~\bibnamefont{{Kajino}}},
  \bibinfo{journal}{\apjl} \textbf{\bibinfo{volume}{625}}, \bibinfo{pages}{L1}
  (\bibinfo{year}{2005}), \eprint{astro-ph/0410142}.

\bibitem[{\citenamefont{{Lewis}}(2004)}]{2004PhRvD..70d3011L}
\bibinfo{author}{\bibfnamefont{A.}~\bibnamefont{{Lewis}}},
  \bibinfo{journal}{\prd} \textbf{\bibinfo{volume}{70}},
  \bibinfo{pages}{043011} (\bibinfo{year}{2004}), \eprint{astro-ph/0406096}.

\bibitem[{\citenamefont{{Plaga}}(1995)}]{1995Natur.374..430P}
\bibinfo{author}{\bibfnamefont{R.}~\bibnamefont{{Plaga}}},
  \bibinfo{journal}{\nat} \textbf{\bibinfo{volume}{374}}, \bibinfo{pages}{430}
  (\bibinfo{year}{1995}).

\bibitem[{\citenamefont{{Bertone} et~al.}(2006)\citenamefont{{Bertone}, {Vogt},
  and {En{\ss}lin}}}]{2006MNRAS.tmp..602B}
\bibinfo{author}{\bibfnamefont{S.}~\bibnamefont{{Bertone}}},
  \bibinfo{author}{\bibfnamefont{C.}~\bibnamefont{{Vogt}}}, \bibnamefont{and}
  \bibinfo{author}{\bibfnamefont{T.}~\bibnamefont{{En{\ss}lin}}},
  \bibinfo{journal}{\mnras} pp. \bibinfo{pages}{602--+} (\bibinfo{year}{2006}),
  \eprint{astro-ph/0604462}.

\bibitem[{\citenamefont{{Razzaque} et~al.}(2004)\citenamefont{{Razzaque},
  {M{\'e}sz{\'a}ros}, and {Zhang}}}]{2004ApJ...613.1072R}
\bibinfo{author}{\bibfnamefont{S.}~\bibnamefont{{Razzaque}}},
  \bibinfo{author}{\bibfnamefont{P.}~\bibnamefont{{M{\'e}sz{\'a}ros}}},
  \bibnamefont{and} \bibinfo{author}{\bibfnamefont{B.}~\bibnamefont{{Zhang}}},
  \bibinfo{journal}{\apj} \textbf{\bibinfo{volume}{613}}, \bibinfo{pages}{1072}
  (\bibinfo{year}{2004}), \eprint{astro-ph/0404076}.

\end{thebibliography}

\end{document}